\documentclass[english,useAMS,usenatbib]{mnras}
\usepackage[T1]{fontenc}
\usepackage[latin9]{inputenc}
\usepackage{graphicx}
\usepackage{amsmath}
\usepackage{xcolor}
\usepackage{url}
\usepackage[authoryear]{natbib}
\usepackage{caption}

\hypersetup{
   unicode=true,            % non-Latin characters in Acrobat's bookmarks
   linktocpage=true,
   pdftoolbar=true,                % show Acrobat's toolbar?
   pdfmenubar=true,              % show Acrobat's menu?
   pdffitwindow=true,           % window fit to page when opened
   pdfstartview={FitH},        % fits the width of the page to the window
   pdfhighlight={/I},
   hyperindex=true,
   pdftitle={Radiometer Design and Calibration for LEDA},            % title
   pdfauthor={Danny C Price et. al.},         % author
   pdfsubject={Subject},       % subject of the document
   pdfcreator={Danny C Price},       % creator of the document
   pdfproducer={Danny C Price}, % producer of the document
   pdfkeywords={keywords},      % list of keywords
   pdfnewwindow=true,           % links in new window
   colorlinks=true,             % false: boxed links; true: colored links
   linkcolor=red,               % color of internal links
   citecolor=blue,              % color of links to bibliography
   urlcolor=cyan                % color of external links
}

\newcommand{\newstuff}[1]{\color{black}{#1}\color{black}}

\renewcommand{\ss}[2] {
 \ensuremath{{#1}_{\rm{#2}}}
}

\makeatletter

\pdfpagewidth=\paperwidth
\pdfpageheight=\paperheight

\title[Radiometer design and calibration for LEDA]{Design and characterization of the Large-aperture Experiment to Detect the Dark Age (LEDA) radiometer systems}

\author[Price et al.] {D.C.~Price$^{1,2,3}$\thanks{E-mail: dancpr@berkeley.edu}, 
L.J.~Greenhill$^{1}$\thanks{E-mail: greenhill@cfa.harvard.edu}, 
A.~Fialkov$^{1}$, 
G.~Bernardi$^{4,5}$, 
H.~Garsden$^{1}$, \newauthor
B.R.~Barsdell$^{6}$,
J.~Kocz$^{1,7}$, 
M.M.~Anderson$^{7}$,
S.A.~Bourke$^{7}$,
J.~Craig$^{8}$, 
M.R.~Dexter$^{2}$,\newauthor
J.~Dowell$^{8}$,   
M.W.~Eastwood$^{7}$,
T.~Eftekhari$^{1,8}$,
S.W.~Ellingson$^{9}$
G.~Hallinan$^{7}$,   \newauthor
J.M.~Hartman$^{7}$,  
R.~Kimberk$^{1}$,
T. Joseph W.~Lazio$^{10}$,
S.~Leiker$^{1}$,
D.~MacMahon$^{2}$,  \newauthor 
R.~Monroe$^{7}$,
F.~Schinzel$^{8,11}$,  
G.B.~Taylor$^{8}$,
E.~Tong$^{1}$,
D.~Werthimer$^{2}$, 
D.P.~Woody$^{7}$\\
$^{1}$Harvard-Smithsonian Center for Astrophysics, 60 Garden Street, Cambridge MA 02138 USA;\\
$^{2}$University of California Berkeley, 501 Campbell Hall, Berkeley CA 94720 USA;\\
$^{3}$Centre for Astrophysics \& Supercomputing, Swinburne University of Technology, PO Box 218, Hawthorn, VIC 3122, Australia;\\
$^{4}$SKA SA, 3rd Floor, The Park, Park Road, Pinelands, 7405, South Africa;\\
$^{5}$Department of Physics and Electronics, Rhodes University, PO Box 94, Grahamstown, 6140, South Africa;\\
$^{6}$NVIDIA Corporation, 2701 San Tomas Expressway, Santa Clara, CA 95050, USA;\\
$^{7}$California Institute of Technology, 1200 E California Blvd, Pasadena, CA 91125 USA;\\
$^{8}$Department of Physics and Astronomy, University of New Mexico, Albuquerque, NM 87131, USA;\\
$^{9}$Bradley Dept. of Electrical \& Computer Engineering, Virginia Tech, Blacksburg, VA 24061 USA;\\
$^{10}$Jet Propulsion Laboratory, California Institute of Technology, 4800 Oak Grove Dr., Pasadena CA 91106 USA \\
$^{11}$National Radio Astronomy Observatory, P.O. Box O, Socorro, NM 87801, USA\\
}

\makeatother

\usepackage{babel}
\begin{document}

%\date{Received 09 25 2017}

\maketitle
\pagerange{\pageref{firstpage}--\pageref{lastpage}} \pubyear{2017}

\label{firstpage}
\begin{abstract}

The Large-Aperture Experiment to Detect the Dark Age (LEDA) was designed to detect the predicted O(100)\,mK sky-averaged absorption of the Cosmic Microwave Background by Hydrogen in the neutral pre- and intergalactic medium just after the cosmological Dark Age.  The spectral signature would be associated with emergence of a diffuse Ly$\alpha$ background from starlight during `Cosmic Dawn'. Recently, \citet{Bowman:2018} have reported detection of this predicted absorption feature, with an unexpectedly large amplitude of 530\,mK, centered at 78\,MHz. Verification of this result by an independent experiment, such as LEDA, is pressing. In this paper, we detail design and characterization of the LEDA radiometer systems, and a first-generation pipeline that instantiates a signal path model. Sited at the Owens Valley Radio Observatory Long Wavelength Array, LEDA systems include the station correlator, five well-separated redundant dual polarization radiometers and backend electronics. The radiometers deliver a 30--85\,MHz band ($16<z<34$) and operate as part of the larger interferometric array, for purposes ultimately of {\it in situ} calibration. Here, we report on the LEDA system design, calibration approach, and progress in characterization as of January 2016. The LEDA systems are currently being modified to improve performance near 78\,MHz in order to verify the purported absorption feature.

\end{abstract}


\section{Introduction}

Cosmic Dawn is a cosmological epoch extending between the build up of the very first population of stars $\sim 100$ Myr after the Big Bang \citep[$z\sim30$][]{Bromm:2011, Greif:2015, Hirano:2016}, followed by corresponding generations of black holes \citep[e.g.,][]{Becerra:2015, Smith:2017, Smidt:2017}, to the onset of widespread reionization of the intergalactic medium (IGM) $\sim 500$ Myr after the Big Bang \citep[$z\sim 10$,][]{Robertson:2015}. This is one of the most interesting and least understood epochs in the history of the Universe \citep[for a recent review see, e.g., ][]{Barkana:2016, Haiman:2016}. Cosmic Dawn is marked by the rise of the earliest populations of sources (stars and black holes), rapid evolution of radiation fields, and the onset of metal enrichment  \citep{Safranek-Shrader:2016, Wise:2014}. 

\newstuff{
Recently, \citet{Bowman:2018} reported detection of the sky-averaged spectral signature of the 21-cm ground-state transition of neutral Hydrogen (HI), placing Cosmic Dawn at redshifts 20>z>15. This signal, predicted by \citet{shaver1999}, is sensitive to both cosmological and astrophysical processes in the early Universe; as such, it is an excellent probe of the physics between the CMB decoupling and the end of the epoch of reionization. Indeed, if verified, the \citet{Bowman:2018} result would constitute the earliest detection of the thermal footprint of the first stars \citep{Greenhill:2018}. 

Specifically, \citet{Bowman:2018} report detection of an $\sim$530\,mK absorption feature, centered at $\sim$78.1\,MHz, with width $\sim$18.7\,MHz, using a relatively simple---yet exquisitely calibrated---dipole antenna and radiometer system known as the Experiment to Detect the Global EoR Step \citep[EDGES,][]{rogers2012,Monsalve:2017a}. The amplitude of this absorption feature is, remarkably, 2--3 times higher than that expected with the most optimistic models \citep{pritchard2010,fialkov2014,Fialkov:2016,Cohen:2016}. Also at odds with existing models, the feature is flat-bottomed, as opposed to Gaussian-like. The \citet{Bowman:2018} result suggests gas temperatures during Cosmic Dawn were far cooler than previously predicted, and could even point toward interaction between baryons and dark-matter particles \citep{Barkana:2018}. An alternative explanation is that there was more radiation than expected, such as a significant contribution from an extragalactic background \citet{Dowell:2018}.

Nevertheless, some concerns remain that the purported Cosmic Dawn signal could in fact be an artifact, due to an unmodelled periodic instrumental feature, for example \citep{Hills:2018}. If verified, the \citet{Bowman:2018} result places virtually the first observational constraints on Cosmic Dawn models. In comparison, the relatively more explored Epoch of Reionization (EoR; $z\sim$ 6--10), is somewhat constrained by (i) the integrated optical depth of Thomson scattering of Cosmic Microwave Background (CMB) radiation \citep{Planck:EoR}, (ii) the high-redshift galaxy UV luminosity function probed out to redshift of $z\sim 10$ \citep{Bouwens:2015, Atek:2015}, (iii) detection of dusty galaxies at redshifts out to $z\sim 10$ \citep{Bouwens:2016, Laporte:2017},  and (iv) supermassive black holes at $z\sim 7$ \citep{Mortlock:2011, Wu:2015}.  

Cosmic Dawn is unique in terms of the astrophysical processes and sources that played roles. In contrast to the EoR, which was likely populated by a `mature' population of galaxies residing in $\sim 10^{8.5}$--$10^{10}$ M$_\odot$ halos \citep{Mesinger2016} and producing copious ionizing radiation, Cosmic Dawn was populated by pockets of intense star formation hosted in dark matter halos of $\sim 10^6$--$10^8$ M$_\odot$, which were less efficient in ionizing their surroundings. Sources of X-rays, Ly$\alpha$ and Lyman-Werner (LW, 11.2--13.6 eV) radiation, on the other hand, played major roles during this epoch \citep[e.g.,][and references therein]{Barkana:2016}, and direct study of this epoch is anticipated to deliver new knowledge about early stellar  populations and to constrain formation scenarios for supermassive black holes (complementary to study of the EoR). 

The preponderance of HI in the diffuse pre- and intergalactic medium (P/IGM) during Cosmic Dawn, and the sensitivity of the transition to radiative backgrounds produced by early stars and black holes makes the 21-cm line a unique tracer of the early Universe.  To date, the main focus of radio instruments undertaking `21-cm cosmology' \citep{pritchard2010}, has been detection of the of EoR power spectrum (i.e. large-scale spatial fluctuations). The Giant Meter-wave Radio Telescope \citep[GMRT,][]{Paciga2013}, the Precision Array for Probing the Epoch of Reionization \citep[PAPER,][]{Ali:2015, Pober:2015}, the Low Frequency Array  \citep[LOFAR,][]{Patil:2017} and the Murchison Widefield Array  \citep[MWA,][]{Beardsley2016,ewall2016} have all placed upper limits on the amplitude of the EoR power spectrum. The upcoming Hydrogen Epoch of Reionization Array \citep[HERA,][]{Deboer:2017}, and the Square Kilometre Array telescope \citep{Koopmans:2015}, also seek to constrain the EoR power spectrum.

Several experiments have been deployed in an attempt to measure the global 21-cm EoR signal. The first constraint on the global 21-cm EoR signal was provided by EDGES \citep{bowman2008,rogers2012}, which excluded reionization more rapid than $\Delta z > 0.06$ with 95\% confidence. The Broadband Instrument for Global HydrOgen ReioNisation Signal \citep[BIGHORNS,][]{sokolowski2015}, and the Shaped Antenna measurement of the background RAdio Spectrum \citep[SARAS,][]{patra2013, Singh:2017} also target the global EoR signal, where results from the latter exclude at 68 to 95\% confidence some parameter combinations that correspond to late heating by X-rays in tandem with rapid reionization.  

EDGES is one of several experiments designed to detect the global 21-cm Cosmic Dawn signal. The SARAS 2 experiment \citep{Singh:2018}, SCI-HI \citep[{\em Sonda Cosmol\'{o}gica de las Islas para la Detecci\'{o}n de Hidr\'{o}geno Neutro},][] {voytek2014} and the related Probing Radio Intensity at high $z$ from Marion (PRIZM) experiment, follow similar methodology and instrumentation approaches. The Dark Ages Radio Explorer (DARE) concept proposes a satellite-based radiometer in lunar orbit, where earth occultation and absence of ionospheric effects are favorable \citep{datta2014, Burns:2017}. 

Here, we detail the Large-Aperture Experiment to Detect the Dark Age instrument \citep[LEDA, see also][]{greenhill2012, bernardi2016}, which observes at frequencies between the  HF (3--30\,MHz) and FM (88--108\,MHz) radio broadcast bands (30<$\nu$<88 MHz, $16<z<34$). LEDA is unique in its embedding of radiometers in a densely interferometric array to enable calibration of radiometric data (in part) with observations of celestial sources (Sec.~\ref{sec:beam-meas}) and to create a ready path for exploration of power spectra estimation for Cosmic Dawn. In this paper, we present the design and characterization of the radiometry system for LEDA.

}
The paper is organized as follows. In Section \ref{Sec:Theory} we provide a broad overview of the physics of Cosmic Dawn, details of the expected 21-cm signal, and outline of experimental requirements needed to observe the Cosmic Dawn signal. We also point out astrophysical scenarios that can be detected or ruled out by LEDA. In Section \ref{Sec:Instrument} we set the stage discussing the site and the architecture of the telescope. LEDA radiometers are discussed in Section \ref{sec:radiometer}. Calibration is discussed in Section \ref{sec:calibration}. In Section \ref{sec:char} we characterize the instrument, including gain linearity, reflection and transmission coefficients, receiver temperature, noise diode thermal stability, and temporal stability. Results, including absolute calibration, RFI occupancy, spectral index measurements, and comparison to extant sky models appear in Section \ref{Sec:results}. Discussion folows in Section \ref{Sec:discussion}. 

\section{Science Driver}
\label{Sec:Theory}

The 21-cm line is a tracer of HI at all stages of cosmic evolution. Before the end of EoR, the signal is mainly produced by the intergalactic neutral medium; at lower redshifts, galactic HI  dominates. The signal is sensitive to both cosmological and astrophysical processes, and as such, is arguably the best probe of the Universe at the intermediate redshift range between the CMB decoupling and the end of reionization.

The observed differential brightness temperature relative to the CMB is
 \begin{align}
  T_{21}  \approx 27 x_{\rm HI}
 \sqrt{\frac{1+z}{10}}\left(\frac{T_S-T_{CMB}}{T_S}\right)\,\textrm{[mK]},
 \label{Eq:T}  
 \end{align}
where we have ignored terms of $\mathcal{O}(\delta)$, with $\delta$  being spatial density fluctuations.  In Eq. (\ref{Eq:T})  $x_{\rm HI}$ is the HI fraction, $T_{CMB}$ is the CMB temperature, and $T_S$ is the spin temperature defined as 
\begin{equation}
T_S^{-1} = \frac{T_{CMB}^{-1}+x_\alpha T_c^{-1}+x_cT_K^{-1}}{1+x_\alpha +x_c}
\label{Eq:S}
\end{equation}
where $T_K$ and $T_c$ are  kinetic and collisional temperatures, and $x_\alpha$ and $x_c$ are Ly$\alpha$ and collisional coupling factors, respectively.
Evolution of $T_S$ with redshift is complex and  depends on the intensity of the Ly$\alpha$ radiative background through  the Wouthuysen-Field (WF) effect \citep{wouthuysen1952, field1958} which sets $x_\alpha$, and on the  thermal history of the Universe via $T_K$, $T_c$ (with  $T_c\approx T_K$) and $x_c$. The temperature $T_{21}$ may be positive or negative  depending on the sign of $\left(T_S - T_{CMB}\right)$. For instance, when the transition is coupled to the gas temperature and the P/IGM is colder than the CMB, $T_S<T_{CMB}$ and the signal is seen in absorption ($T_{21}<0$).  When the gas is hotter than the CMB, $T_S>T_{CMB}$ and the signal is seen in emission ($T_{21}>0$).   

\begin{figure}
\includegraphics[width=1\columnwidth]{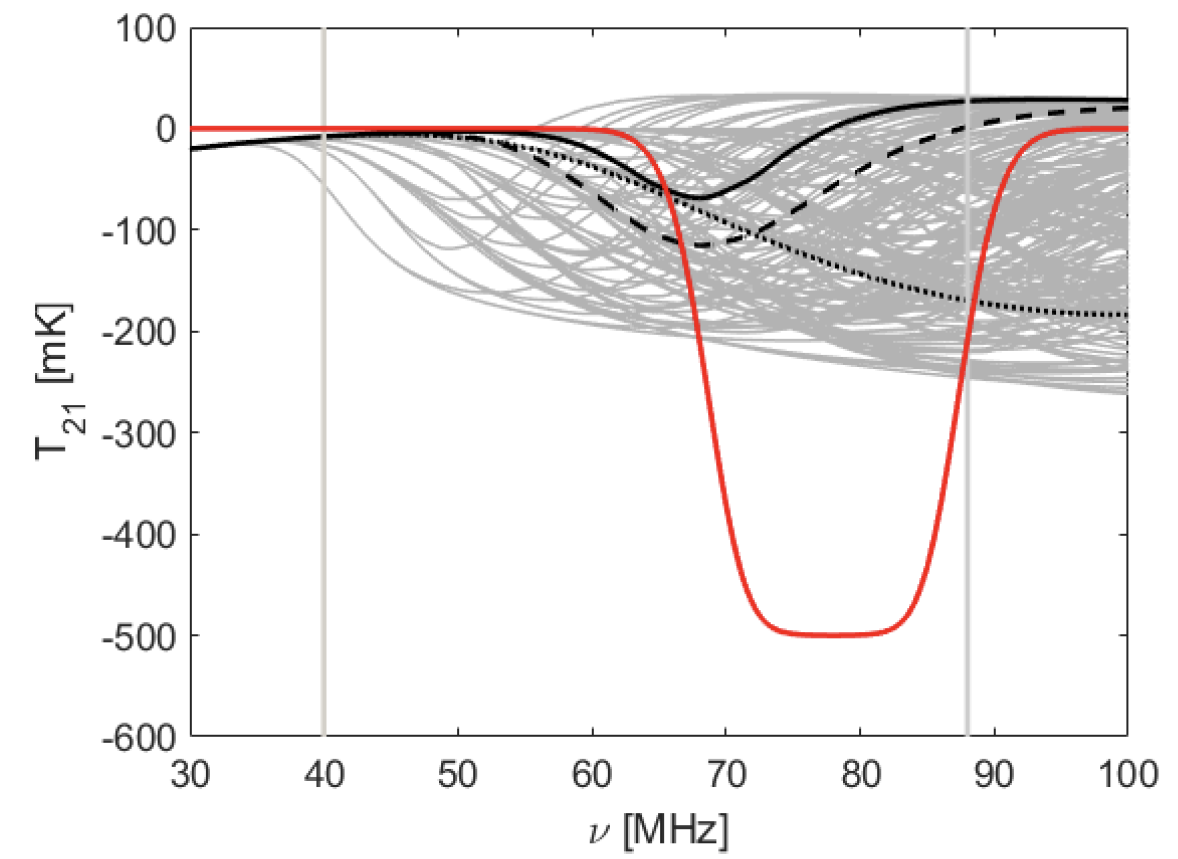}
\protect\caption{\label{fig:models} \newstuff{Spectra for variations in model parameters taken from \citet{Cohen:2016}, with purported EDGES detection from \citet{Bowman:2018} in red. The heavy black traces show three particular scenarios with strong (solid), intermediate (dashed) and weak (dotted)  signatures in the LEDA observing band (delineated by vertical solid lines).}}
\end{figure}

For any particular scenario of structure and star formation,  the evolution of $T_{21}$ can be used as a `cosmic clock' that tracks the evolution of the Universe. In what follows, we focus on the zero-mode of the signal  from  Eq. \ref{Eq:T},  a.k.a.,  the global signal\footnote{Higher order modes are linked to  spatial fluctuations and are outside of the scope of this paper \citep[for more details see][and references therein]{Barkana:2016}.}. The predominant feature of the signal (e.g., the black curve in Fig. \ref{fig:models} which is a model with a strong signature in the LEDA observing band) is a Gaussian-like absorption trough signifying  sufficient Ly$\alpha$ coupling and a cold diffuse medium. The centroid and amplitude of the trough depend directly on the balance between the processes of star formation and P/IGM heating. In particular, the low-frequency maximum  (located at $\nu = 46$ MHz for the black curve in Fig. \ref{fig:models}) denotes the onset of star formation giving rise to the Ly$\alpha$ background. These photons coupled $T_S$ to $T_K$ via the WF effect, creating an absorption signature because at that epoch  gas was colder than the CMB.  The strongest absorption  (at $\nu = 68$ MHz for the black curve in Fig. \ref{fig:models}) marks approximately the moment at which the IGM has reached its minimum temperature and a growing X-ray background due to compact sources becomes significant.  As cosmic heating progresses, contrast between $T_S$ and $T_{CMB}$ decreases until  the moment when the gas reaches the temperature of the background radiation, and the signal vanishes (at $\nu = 87$ MHz for the black curve in Fig. \ref{fig:models}).   If this happens prior to reionization by ultraviolet sources, the signal will appear in   emission for lower $z$. 

The 21-cm spectra for an ensemble of  astrophysical model parameters combinations ($M_{min}$, $f_*$, $L_X$, X-ray SED and the total CMB optical depth $\tau$)  permitted by extant data and theoretical studies exhibit a large scatter  as is shown in Fig. \ref{fig:models}\newstuff{;  the recent \citet{Bowman:2018} result is overlaid in red}. The amplitude of the CD trough varies between 25 and 240\,mK, with the absorption trough located between 40<$\nu$<120 MHz \citep{Cohen:2016}.
\newstuff{The \citet{Bowman:2018} result is inconsistent with these models, exhibiting a much larger amplitude of 530\,mK. Given this discrepancy, coupled with some outstanding concerns that the signal is an artifact \citep{Hills:2018}, and that the result is yet to be verified, we do not as of yet rule these scenarios out.}  

For a large fraction of physically motivated models, \newstuff{and also for the purported \citet{Bowman:2018} signal}, the absorption minimum falls within the LEDA observing band, and, thus, could be detected by the instrument. This is discussed further in the next section.

\subsection{Observational Prospects}
\label{sec:prospects}

Radiometric detection requires separation of foreground signals and the (background) 21-cm  signal. The diffuse and continuum foreground sources are know to be spectrally smooth; that is, they exhibit power-law spectra over the 30--88\,MHz band. As such, they are separable from the background signal, which is expected to manifest as an absorption trough. \citet{pritchard2010} showed a basic demonstration of concept, by convolving a Global Sky Model \citep[GSM,][]{deOliveiraCosta:2008ks} with an analytical model for a simple dipole antenna to form simulated measurements; this approach is followed by several Cosmic Dawn experiments \citep{Singh:2018, voytek2014, Bowman:2018}, including LEDA.

 The spectral smoothness of foregrounds allows retrieval of 21-cm features by modeling the brightness temperature of the foreground, $T_{fg}$, with a low-order log-polynomial. \cite{harker2012} expounded on this by including instrumental effects; \cite{bernardi2015} showed that the angular structure and frequency dependence of a more realistic broadband dipole (modeled on the design of the Long Wavelength Array) increases the required polynomial order but not necessarily so much so as to confound detection. 
For this approach to work, any spectral structure introduced by the measurement apparatus must be accounted for and calibrated out. For this reason, zero-mode radiometer experiments have preferred simple, low-gain dipole antennas over high-gain single dishes that exhibit more complex gain patterns \citep{bernardi2015, mozdzen2016}. We discuss the approach of LEDA to calibration  in Section \ref{sec:calibration}.

Significance of the detection by LEDA is determined by tiny deviations in the shape of the actual sky temperature from the smooth foreground curve in the LEDA band. To estimate which part of the astrophysical parameter space is actually targeted by LEDA, we use the signal to noise ratio (SNR) defined as 
\begin{equation}
SN^2 = \sum_i \frac{(T_{sky}-\tilde T_{fg})^2}{\sigma_i^2}
\end{equation} 
where the sum is over frequency channels in the LEDA band, the mock data is defined as $T_{sky} = T_{fg} + T_{\rm{cosm}}$ and the  foreground signal is modeled as a seventh order polynomial in $\log\nu$   \citep{bernardi2016}. We fit out the foreground component by calculating $\tilde T_{fg}$ as a best fit  to the mock data of the shape provided by \citet{bernardi2016}. The residual signal is then compared to the rms noise given by the radiometer equation 
\begin{equation} 
 \sigma \approx 2.6\times\frac{T_{sys}}{\sqrt{\Delta\nu t}}
 \end{equation}
where  $\Delta\nu$ is the bandwidth over which the signal is measured, and $t$ is the integration time, and we assumed that the system temperature, $T_{sys}$, is dominated by the temperature of the sky. The factor 2.6 is an approximation that takes into account thermal uncertainties after calibration. (Here we assumed hot and cold reference diode temperatures of 6500\,K and 1000\,K, as is explained in detail in Sec.~\ref{sec:calibration}, Eq.~\ref{eq:rms-unc}). \newstuff{Thus for a sky temperature of 2000\,K and using $\Delta\nu$=1\,MHz, a $5\sigma$ detection of the \citet{Bowman:2018} feature could be made in under 45\,minutes of observation with LEDA.}

\newstuff{Alternatively, one may } calculate SNR for each of the $\sim 200$ models shown in  Fig. \ref{fig:models}, assuming $\Delta \nu$ = 1\,MHz and integration time of 1000 hours. Out of $\sim 200$ different astrophysical scenarios (Fig. \ref{fig:models}) the models with the highest SNR are those with strongest variation within the LEDA band.  These models typically share high star formation efficiency (often in low-mass halos) and high X-ray efficiency, which suggests that LEDA should have considerable leverage  in constraining (i) star formation during cosmic dawn, and in particular the roles of small halos, and (ii) the timing of X-ray heating and properties of high-redshift X-ray sources (e.g., XRB, mini-quasars). The model with the highest signal to noise of SNR = 9.2 (heavy solid curve in Fig. \ref{fig:models}) shows both a strong absorption and an early emission signal within the LEDA band. These features are hard to mimic with smooth foregrounds. The underlying astrophysical model assumes high star formation efficiency of $f_* = 50\%$ in heavy halos above circular velocity of 35.5 km s$^{-1}$ and a very luminous XRB population  shining at the luminosity of  $L_X = 15 \times 10^{41}$ erg s$^{-1}$ per unit star formation rate in M$_\odot$ yr$^{-1}$ (i.e., 50 times brighter than the low-redshift counterparts).  

In another detectable scenario (heavy dashed curve) only the absorption trough is located within the LEDA band which makes the detection a bit more challenging. The underlying astrophysical model has moderate star formation efficiency, $f_* = 5\%$,  stars form via cooling of atomic hydrogen, X-ray heating is due to XRB with $L_X = 2.4 \times 10^{41}$ erg s$^{-1}$ per unit star formation rate in M$_\odot$ yr$^{-1}$, and the total CMB optical depth of $\tau = 0.066$. The moderate star formation and heating result in a moderate SNR = 4 in the LEDA band. Finally, in Fig. \ref{fig:models} we also show an astrophysical scenario that cannot be separated from the foregrounds, and, thus, is  undetectable by LEDA (SNR = 0.1, dotted line in the figure). This case has low star formation efficiency, $f_* = 0.5\%$, star formation via molecular hydrogen cooling subjected to strong LW feedback, and weak X-ray heating with $L_X = 0.03 \times 10^{41}$ erg s$^{-1}$. 

\section{Instrument overview}
\label{Sec:Instrument}

Motivated by limitations of single-antenna experiments, LEDA is a multi-antenna experiment, co-installed on the Long Wavelength Array stations at Owens Valley Radio Observatory (OVRO-LWA, 37.24$^\circ$N, 118.28$^\circ$W), and the National Radio Astronomy Observatory in Socorro, New Mexico (LWA1, 34.07$^\circ$N, 107.63$^\circ$W). Initial work, based at LWA1, is presented in (Schinzel et al., LWA Memo \#208); here we focus on work carried out at OVRO-LWA. Further details about the LWA systems may be found in \citet{taylor2012,ellingson2013}.

\begin{figure}
\includegraphics[width=1\columnwidth]{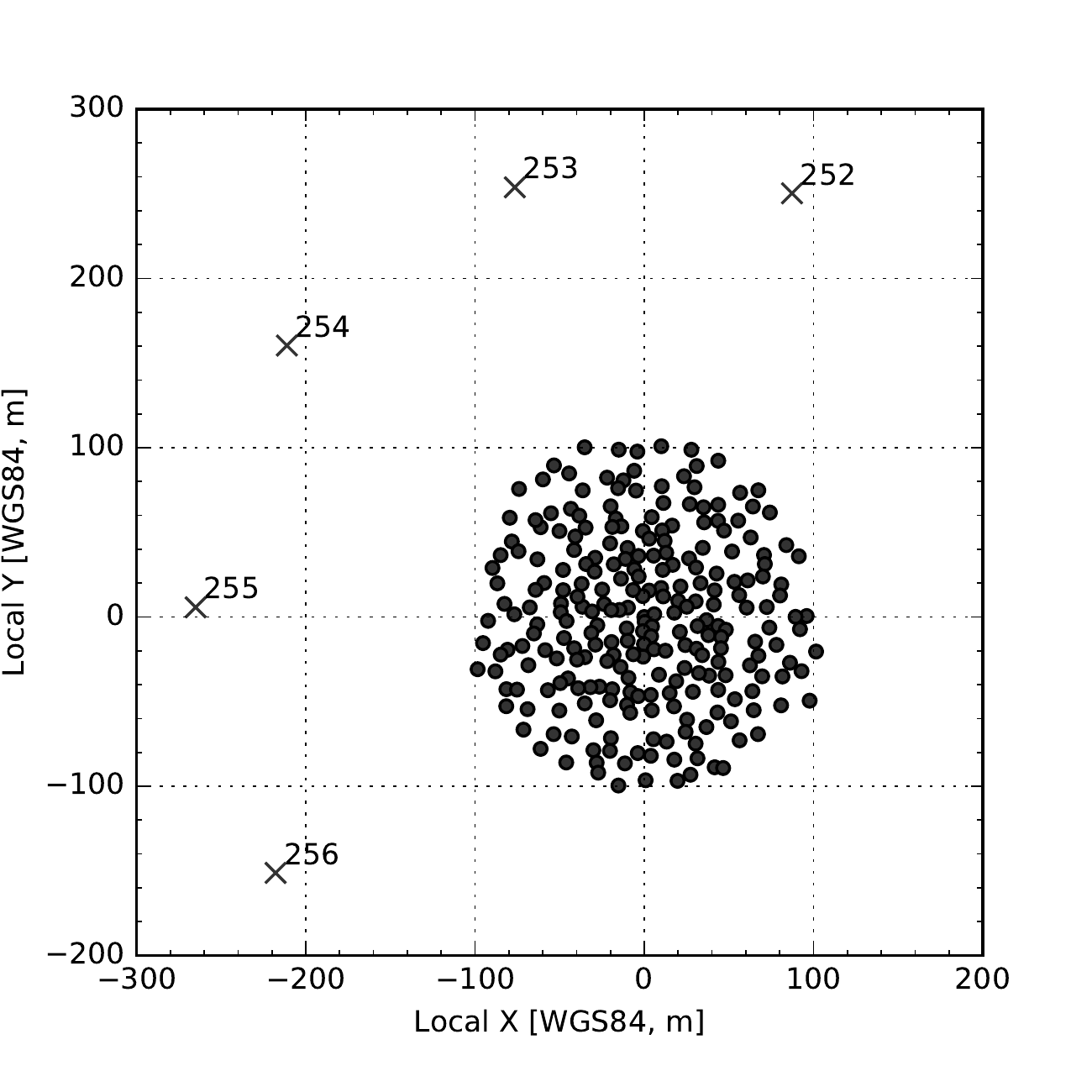}
\protect\caption{\label{fig:ant-pos} Antenna positions in the  WGS84 coordinates relative to the mean.
Crosses represent the outrigger antennas equipped with LEDA radiometric frontends. (Antennas installed at greater separations in 2015 are not shown.}
\end{figure}

At both the LWA1 and OVRO-LWA stations, an additional 5 outrigger stands (Fig.~\ref{fig:ant-pos}) were installed and outfitted with the LEDA frontend receiver card (see Sec.~\ref{sec:radiometer}). The LEDA outrigger stands, detailed further in Sec.~\ref{sec:outriggers}, are placed at a distance from the core to minimize mutual coupling effects. \citet{bernardi2015} argue that detection will require precise knowledge of the antenna radiation pattern, which may not be deliverable by EM simulation. \citet{datta2014} are pessimistic that a detection may be made without an accurate model of the ionosphere. There is also concern that terrestrial radio frequency interference (RFI) could confound detection. 

\begin{figure*}
\includegraphics[width=2\columnwidth]{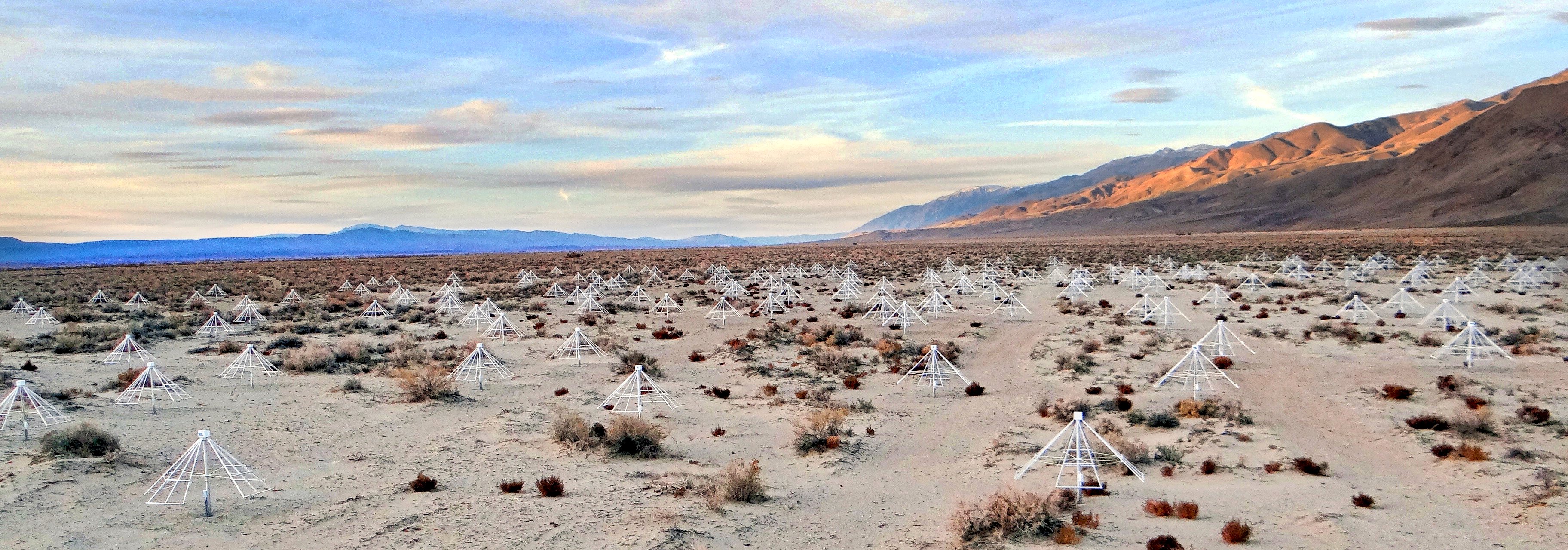}
\protect\caption{\label{fig:lwaphoto}The core area of the Owens Valley LWA, looking north from the electronics shelter located at the edge.  The dual-polarization dipoles are aligned  north-south and east-west. }
\end{figure*}

The OVRO-LWA core region consists of 251 stands within a 100-m radius  (Fig~\ref{fig:ant-pos}), roughly double the radius of LWA1. 
The OVRO-LWA station (Fig.~\ref{fig:lwaphoto}) was built in 2013 and utilizes the same antenna design and analog systems as LWA1, with a different digital system designed for wide-bandwidth cross-correlation \citep{kocz2015}. An additional 5 outrigger stands (Fig~\ref{fig:ant}) were installed and outfitted with the LEDA frontend receiver card---the characterization and design of which is the focus of this article. A more complete overview of OVRO-LWA may be found in Hallinan et al., in prep.  

%%%% FIGURE

\subsection{Observational strategy}

OVRO-LWA allows for LEDA to monitor the ionosphere, characterize the foreground sky, and measure antenna gain patterns {\em in-situ}, all while radiometric measurements are being taken. Gain patterns are measured via cross-correlation of LEDA radiometer antennas against the dense core of OVRO-LWA. The OVRO-LWA array is capable of imaging the radio sky from horizon to horizon; by doing so, the position and apparent brightness of celestial sources can be monitored. By monitoring the passage of celestial sources, a beam model can be inferred; similarly, a model of ionospheric-induced refractive offsets may be formed by monitoring offsets in the apparent positions of sources.

The LEDA observational strategy is implemented to avoid complications that arise due to synchronization requirements between the noise diode switching and correlator integration. In order to derive gain patterns, once a week a 24-hour interferometric observation of the sky is performed at a low duty cycle (a 10\,s integration every 100\,s). These data are also used to form a model of the sky. During these observations, the radiometer antennas do not switch into diode states. In contrast, during radiometric observations the ionosphere is monitored using the core antennas and cross-correlations with the outriggers are discarded.

The primary LEDA observational window is during night-time hours over December-March. The Sun is a potential source of interference, and RFI is known to be more prevalent during daylight hours. Furthermore, at low frequencies the galactic plane has a significant contribution to the overall system temperature of a radiometer, so it is desirable to observe when the galactic plane is low.

%%%% FIGURE

\subsection{LEDA correlator and spectrometer} \label{sec:corr}

OVRO-LWA is operated as a radio interferometer, which requires the cross correlation of all antenna pairs. Cross correlation is an $O(N^2)$ operation, which is computationally challenging for the $N=$512 inputs of OVRO-LWA. The cross-correlation of all 256 dual-polarization antenna pairs is performed by the LEDA correlator, detailed in \citet{kocz2014b, kocz2015}. Briefly, the LEDA correlator is a FX-style system where the data are channelized (by `F-engines') before cross-correlation by the `X-engine'. The F-engines run on ROACH-2 field-programmable gate array  (FPGA) boards from the CASPER collaboration \citep{Hickish:2016}. The F-engines are connected via 10\,Gb Ethernet to compute servers running the xGPU cross-correlation X-engine code \citep{Clark:2013}. 

The firmware for the LEDA F-engine has been modified from that detailed in \citet{kocz2015} to generate autocorrelation spectra at higher bit depth. The LEDA correlator requantizes the output of the polyphase filterbank down to 4 bits, for data transport to the compute servers. The modified firmware generates autocorrelation spectra from the (18-bit) F-engine output before requantization, yielding higher dynamic range and lowering quantization-induced non-linearity compared to the 4-bit data stream. We refer to this system as the LEDA spectrometer.

All spectral data products presented here are from the LEDA spectrometer. Corresponding cross-correlation data from the LWA core were also recorded for ionospheric monitoring, but are not used further within this article. The spectrometer is implemented using a 4096-channel, 4-tap, Hamming-windowed polyphase filterbank (PFB). The PFB provides $\sim$50\,dB of isolation between neighboring channels, which prevents leakage of narrowband RFI signals between channels. The digitizer is clocked at 196.608\,MHz, resulting in a 24\,kHz channel bandwidth. The output of each PFB channel is squared and accumulated for 1\,s. An external pulse-per-second (PPS) signal is used to trigger each new accumulation. Accumulated data are read from the ROACH-2 board's Ethernet control interface. After every accumulation, data are timestamped and written to a hierarchical data format (HDF5) file. 

\subsection{Outrigger antennas} \label{sec:outriggers}

%%%% FIGURE
\begin{figure}
  \includegraphics[width=0.99\columnwidth]{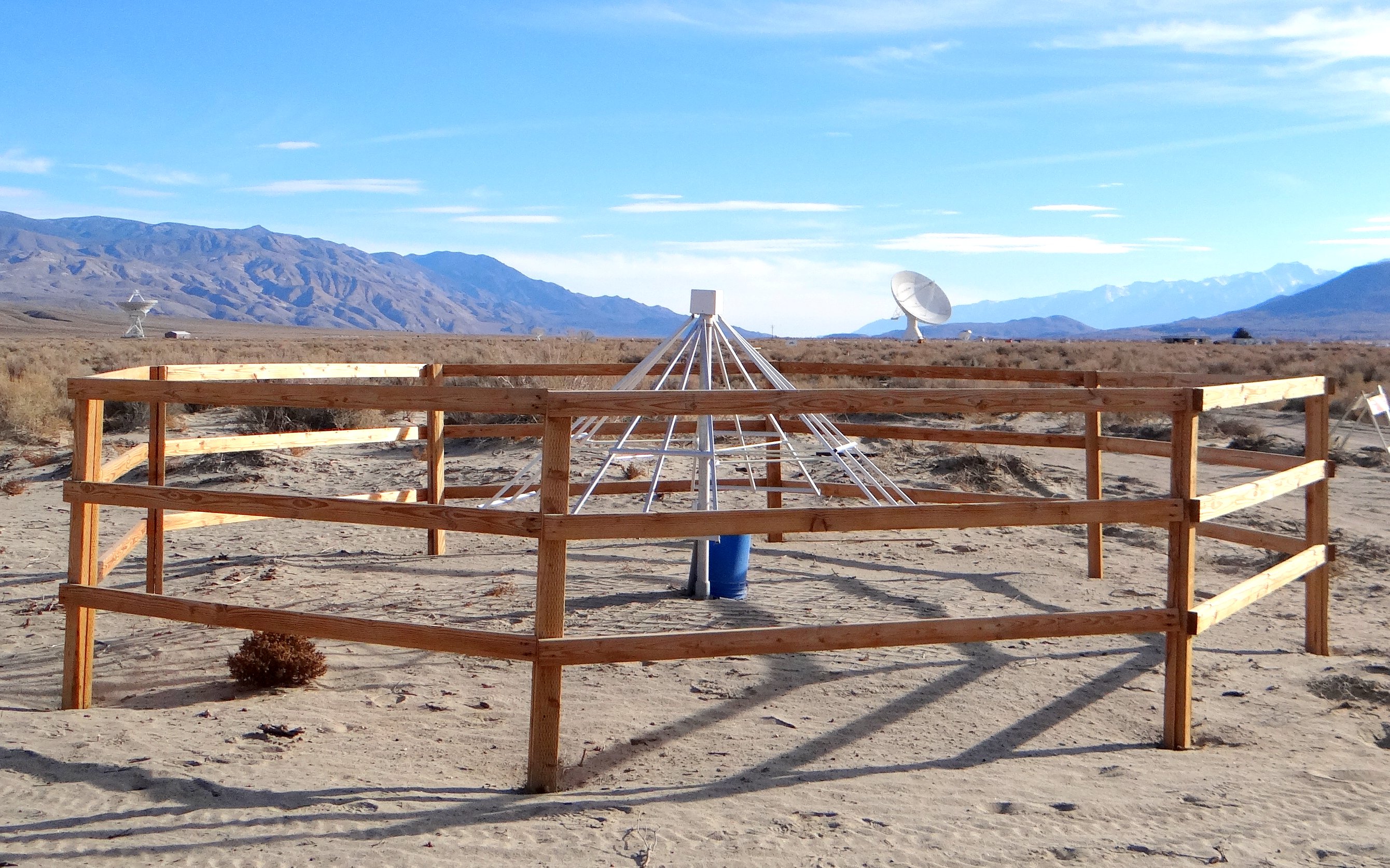}
  \caption{LEDA outrigger antenna stand (ID 256) at Owens Valley Radio Observatory, California, January 2015. \label{fig:ant}}
\end{figure}

The outrigger antennas (Fig.~\ref{fig:ant}) are of the same design as the LWA cross-dipole antennas, detailed in \citet{ellingson2013}. Each dual-polarization antenna consists of four triangular 'blades' of length 1.4\,m, which form two pairs of orthogonal antennas (single-polarization). The blades are attached to a central pole of height 1.5\,m, at the top of which is a weatherproof box for the FE. The two antenna pairs are oriented North-South and East-West. The blades are angled down to improve response at the horizon and beam symmetry. A wire-grid 3$\times$3-m ground screen isolates the antennas from the earth ground, whose characteristics may change with moisture content.

The outrigger antennas are physically isolated from the core antennas and other metallic objects. Each antenna has a 3$\times$3\,m ground screen, and is protected from grazing cattle by a wooden fence. The antennas are connected to the shelter via buried lengths of LMR400 coaxial cable. These cables are fed up through the central pole of the antenna, through to the receiver.

\section{LEDA Radiometers}
\label{sec:radiometer}

\begin{figure*}
\includegraphics[width=1.9\columnwidth]{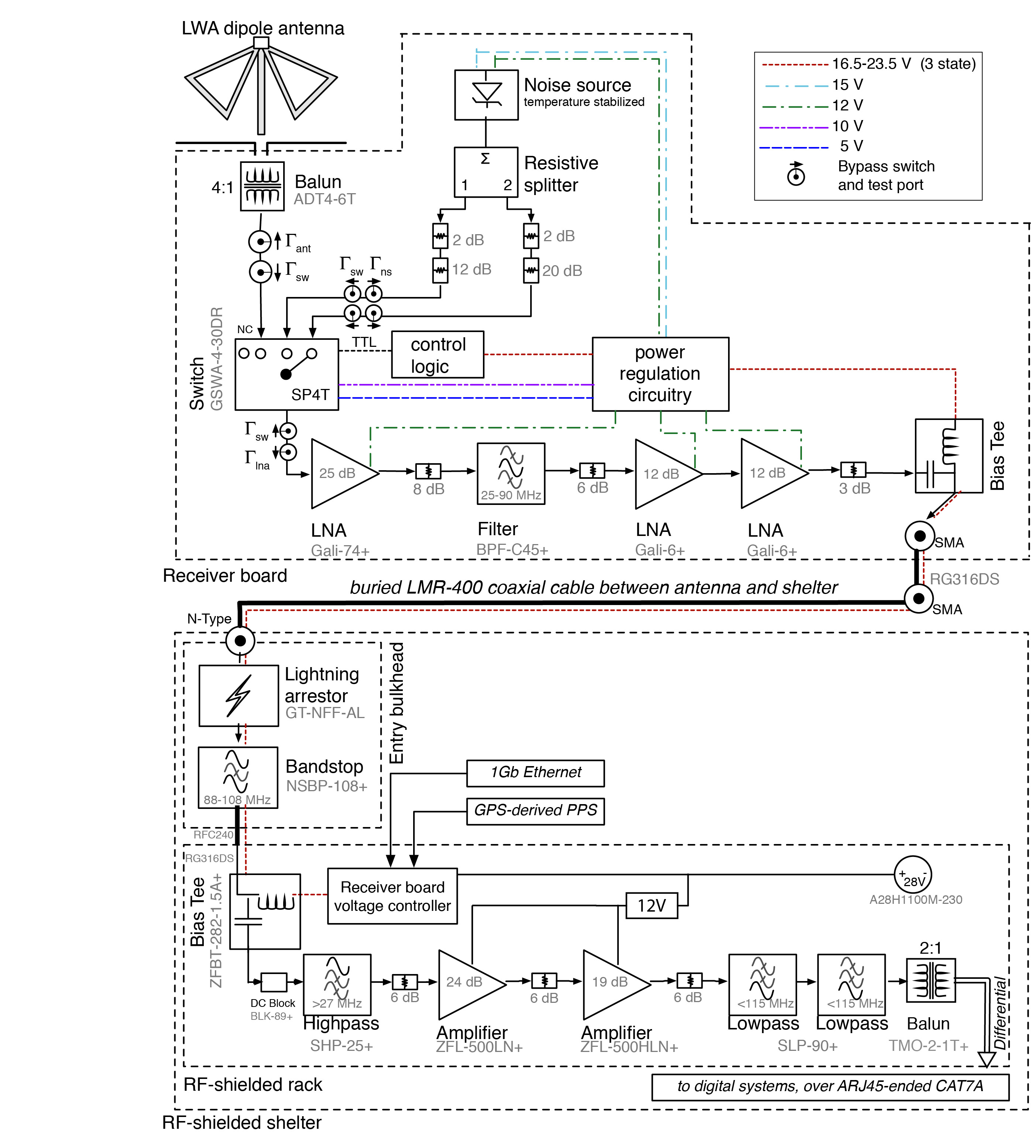}
\protect\caption{Block diagram of a single polarization radiometer analog signal path.  Components on a receiver board are shown in the dashed box in the upper half.  (See also Figure\,\ref{fig:fe}.)  RF jacks (Hirose MS147) provide test ports for measurement of reflection coefficients.  When in use, the through path is broken, and the network analyzer looks in the direction indicated.  Components in the central electronics shelter are in the lower half.  Details of the temperature stabilized noise source are shown in Figure\,\ref{fig:diode}.  DC supply lines are in color.   
\label{fig:block_diag} }
\end{figure*}

In this section we provide further detail about a single radiometer system; in total, there are 5$\times$2 complete radiometers within LEDA. A block diagram of a LEDA radiometer is given in Fig.~\ref{fig:block_diag} for a single polarization. As seen in the diagram, there are four main components: the antenna, front-end receiver, back-end analog systems, and digital systems. Each dual-polarization antenna is connected directly at its terminals to the front-end electronics (FE). The FE converts the balanced antenna terminal pair to unbalanced 50\,$\Omega$ via a 4:1 balun, then amplifies and filters the signal, outputting the conditioned signal over a buried coaxial cable to the OVRO-LWA electronics shelter. 

The entry bulkhead of the shelter connects the buried coaxial to an FM bandstop and lightning arrestor installed at the bulkhead; a length of coaxial cable connects the bulkhead to the back-end analog systems (CRX).The CRX applies further amplification and signal filtering, in preparation for digitization. The CRX systems also provide power to the FE, via the coaxial cable lengths that connect the antenna to the shelter. The signal from the CRX is converted from 50\,$\Omega$ unbalanced to 100\,$\Omega$ balanced, and Category-7A Ethernet cables are used to transport the signals to the digitizer. Further details of these systems are given below.

%%%% FIGURE

\subsection{Deployment history} \label{sec:dev_cycle}

The LEDA radiometer systems are under active development; iterative upgrades and improvements are made after each field deployment (Tab.~\ref{tab:dev_cycle}). A brief historical summary of deployments is as follows. The LEDA correlator system achieved first light in August 2013, using an early revision of the FE card (ver 2.0) along with the standard LWA analog receiver system (ARX), with spectra formed from 4-bit data sent to the correlator. In December 2013, the correlator F-engine firmware was modified to add an independent autocorrelator spectrometer with higher bit-depth. An updated version of the FE with added MS147 test ports (ver 2.5) was installed in April 2014. Due to concerns of potential crosstalk, a fully shielded switching controller (Sec.~\ref{sec:analog}) was installed in November 2014, and the LWA analog receiver was replaced by a fully connectorized system in December 2014. Major improvements were made to the FE over the course of 2015; the FE version 2.9 was deployed in January 2016.

 In this paper, we detail the system as installed at OVRO-LWA in January 2016; details of the FE version 2.5 and LWA1 results are presented in Schinzel et. al. (in prep). 

\begin{table}
\caption{Summary of radiometer system upgrades 2013--2016.
	\label{tab:dev_cycle}}
	\begin{tabular}{l l}
\hline 
mmyy      & Deployment milestones  \\%&  Driver   \\ 
\hline
\hline 
08/13  & First light$^{(1)}$                     \\
12/13  & Embedded standalone 8-bit spectrometers$^{(2)}$ \\
04/14  & FE ver. 2.5: bandpass filtering         \\
11/14  & Shielded programmable switching controller      \\
12/14  & Shielded, low cross-talk backend systems, high-\\
       & isolation standard for multi-channel cables$^{(3)}$ \\
01/16  & FE ver. 2.9: improved impedance matches, better \\
       & noise source stability, high RF directivity   \\           
\hline 
\end{tabular} 
\\
\raggedright $^{(1)}$ FE ver. 2.0, unshielded 1\,Hz switching controller, LWA analog backend, and LEDA correlator.\\
\raggedright $^{(2)}$ Phased out use of station correlator for precision radiometric data.  \\
\raggedright $^{(3)}$ Adoption of Bel-Stewart ARJ45 differential cable-end standard for RF over twisted pair between analog receivers and correlator digital samplers.

\end{table}

\subsection{Front-end receiver board}

%%%% FIGURE
\begin{figure}
\includegraphics[width=1.05\columnwidth]{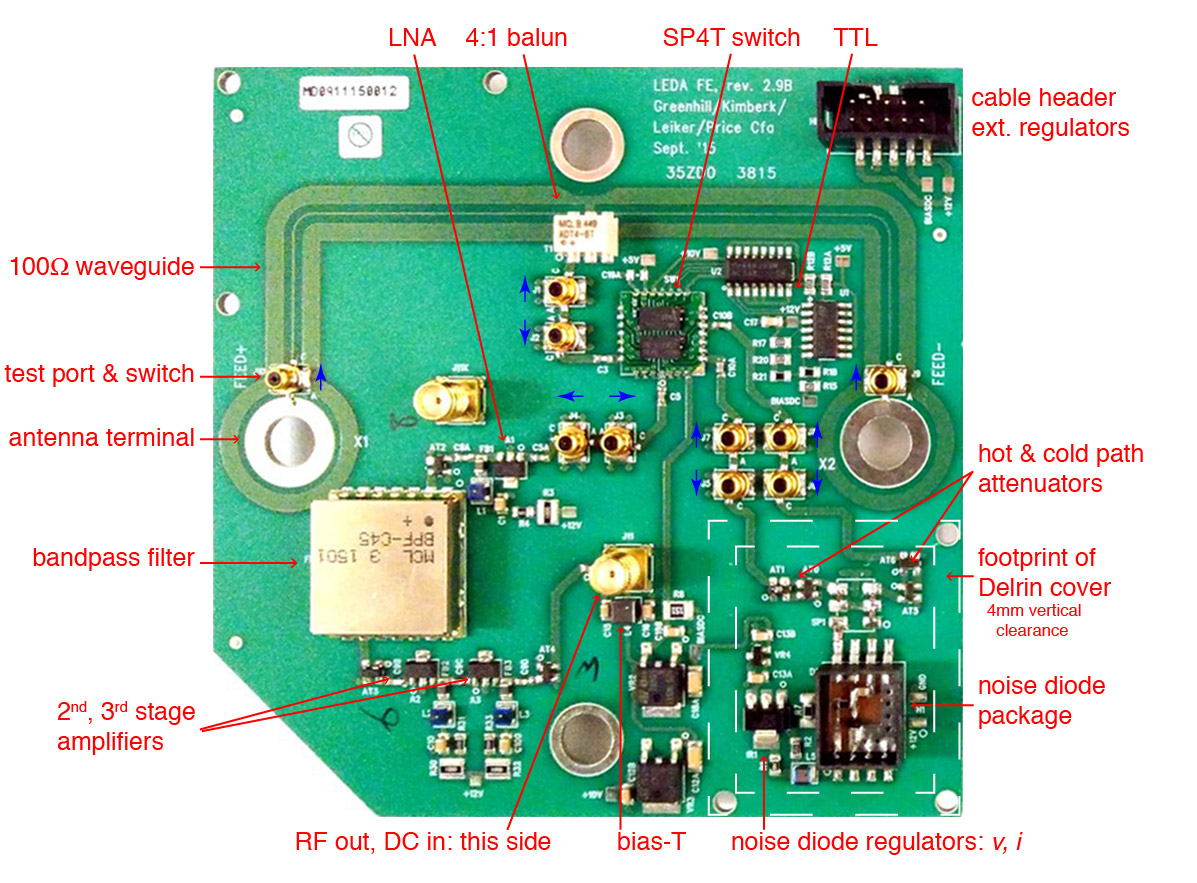}
\protect\caption{Photograph of the LEDA v2.9 FE,
with key components labeled. The Delrin cover and heater has been removed so
that the noise diode is visible. \label{fig:fe} }
\end{figure}

%%%% FIGURE
\begin{figure*}
\includegraphics[width=1.7\columnwidth]{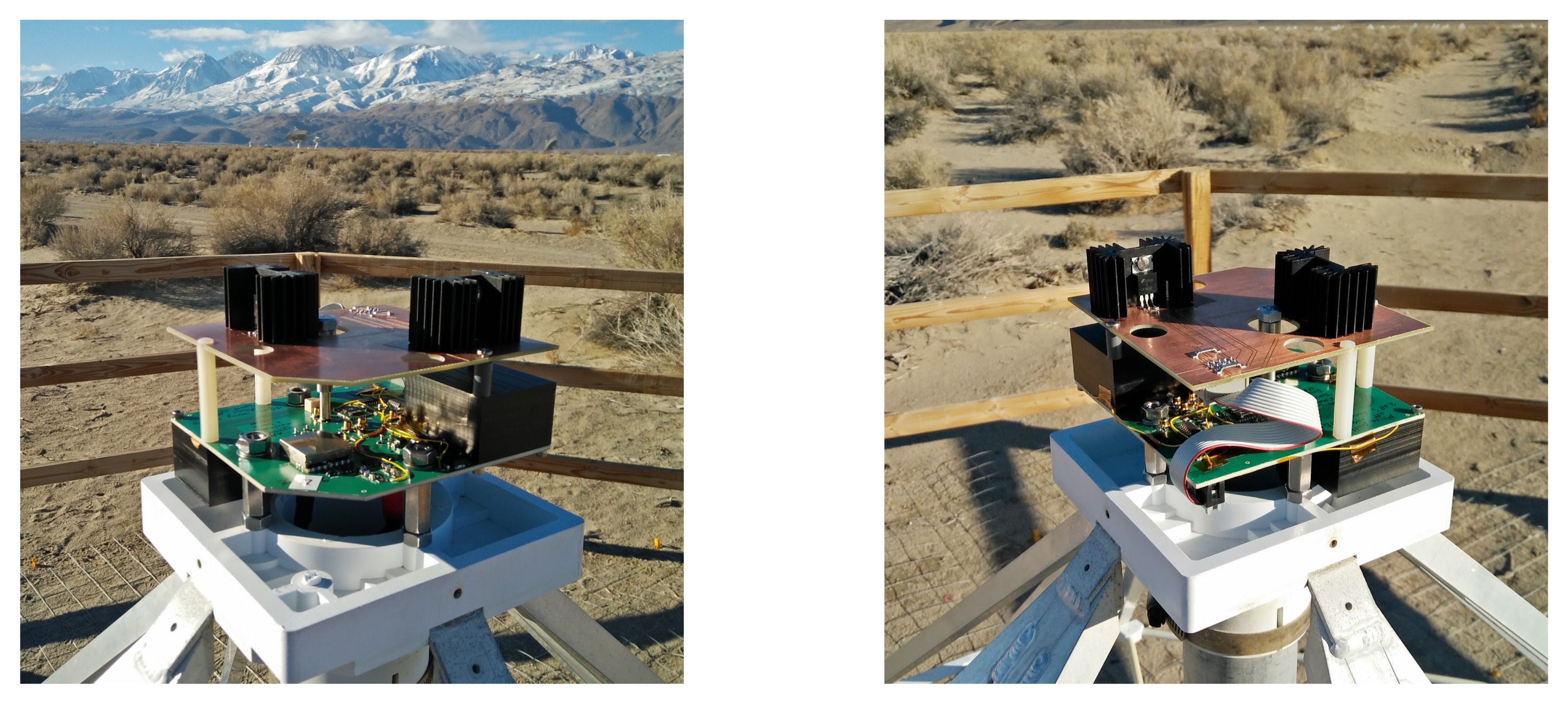}
\protect\caption{LEDA v2.9 radiometer FE, as installed on an OVRO-LWA outrigger antenna, as seen from the Northwest (\textit{left}) and Southeast (\textit{right}).  As for LWA hardware,  the threaded posts serve as antenna terminals.  Upward facing electronics on the FE card serves the east-west dipole.  Downward facing electronics serve the north-south dipole.  For this photo, the protective plastic cover has been removed.\label{fig:fe_installed} }

\end{figure*}

\begin{table*}
	
	\centering
    \caption{Cascaded FE Noise Temperatures at 50\,MHz\label{tab:tsys}}
	\begin{tabular}{l | c c | c c | c c c }
	\hline
	Component & \multicolumn{2}{c}{Noise Figure} & \multicolumn{2}{c}{Gain ($S_{21}$)} & $T_{noise}$ & $\ss{T}{rx}$ & $\ss{G}{rx}$  \\
	          &    (dB)      &(dB\,K$^{-1}$)& (dB) &(dB\,K$^{-1}$)& (K) & (K)          &    (dB)       \\ 
	\hline
	\hline

	GSWA-4-30DR$^{(1)}$ & -- & -- & -0.7 & -- &  53.1   & 54.6   & -0.7  \\
    Gali-74$^{(2,3)}$ & 2.64 & -0.0052 & 24.8  & -0.0026 & 249.4 & 373.7  & 24.1  \\
    LAT8              & -- & -- & -8.1 & -- & 1613.6 & 380.5  & 16.0  \\
    BPF-C45+          & -- & -- & -0.4 & -- & 25.8   & 381.2  & 15.7  \\
    LAT6              & -- & -- & -6.2 & -- & 950.5  & 409.1  & 9.4   \\
    Gali-6$^{(2,3)}$  & 4.27 & -0.0087 & 13.0  & $-0.0015$ & 470.5 & 428.5  & 22.5  \\
    Gali-6$^{(2,3)}$  & 4.27 & -0.0087 & 13.0  & $-0.0015$ & 473.5 & 431.9  & 35.5  \\
    LAT3              & -- & -- & -2.9 & -- & 281.9 & 432.0  & 32.6  \\
	\hline
    \multicolumn{8}{p{12cm}}{$^{(1)}$ Manufacturer specifications at 64 MHz.} \\
    \multicolumn{8}{p{12cm}}{$^{(2)}$ Manufacturer specified gain for T=25\,C, $50\Omega$ source and load, and $i_{cc}=65$\,mA (Gali-74) or $i_{cc}=70$\,mA (Gali-6). Measured currents are 61 and 68\,mA, respectively. Component operating temperatures are $\sim 40$\,C for an ambient temperature of 25C.} \\
    \multicolumn{8}{p{12cm}}{$^{(3)}$ Manufacturer specified noise figure for test conditions in note (2).} \\
\end{tabular}
\end{table*}

The LEDA FE (Fig.~\ref{fig:fe}) connects to the antenna terminals (Fig.~\ref{fig:fe_installed}) and applies first-stage signal amplification and conditioning. The FE is a two-sided, four-layer circuit board of dimensions of 11.5$\times$11.5\,cm, installed in the weatherproof box at the antenna's apex. The FE provides signal paths for both stand polarizations, one per side. 

A differential 200\,$\Omega$ line connects each blade pair to a Mini-Circuits ADT4-6T transformer (balun), that converts the balanced signal to 50\,$\Omega$ unbalanced. A four-throw switch (Mini-Circuits GSWA-4-30DR) allows for selection between the antenna path and two calibration reference paths (Sec.~\ref{sub:cal}). The switch output then connects to the first-stage low noise amplifier (LNA, Mini-Circuits Gali-74+, Tab.~\ref{tab:tsys}). A bandpass filter (Mini-Circuits BPF-C45+) suppresses signals outside of 25--90\,MHz, attenuating HF and FM RFI sources outside the band of interest. Mini-Circuits LAT attenuators are used between components to improve impedance matching.

Second-stage signal amplification is done after filtering, using two Mini-Circuits Gali-6+ amplifiers connected in cascade, each with 12.2\,dB gain. While the Gali-74+ has a lower noise figure than the Gali-6+, the Gali-6+ was chosen for its flatter gain response over the LEDA band. Along the signal path, several buffering attenuators are installed to improve impedance match between components. 

Tab.~\ref{tab:tsys} shows the cascaded gain ($\ss{G}{rx}$) and receiver temperature ($\ss{T}{rx}$) after each component in the FE analog path. The overall receiver temperate $\ss{T}{rx}$=432\,K, and the receiver gain $\ss{G}{rx}$=31.2\,dB. Note that these values are calculated from specifications provided in component data sheets; actual measurements are provided in Sec.~\ref{sec:char}. 

Losses before the first-stage LNA and the LNA's noise temperature dominate $\ss{T}{rx}$. The overall $\ss{T}{sys}$ of a LEDA radiometer is nonetheless dominated by the antenna temperature, which is >1000\,K across the LEDA band.

\subsubsection{DC power and state control}
The FE receives DC power via its SMA output jack. The RF signal and DC power are separated by an on-board bias tee. The switch state is controlled by changing the DC voltage supplied to the board, between 17\,V (sky), 20\,V (cold reference) and 23\,V (hot reference). This allows FE state to be controlled remotely from the electronics shelter.

Many of the components on the board require 12\,V or lower. Regulation to 12\,V is conducted on an external board, which connects to the FE via a ribbon cable; the thermal load of the regulators would produce undesired thermal gradients if placed directly on the FE.

\subsubsection{Test ports}

The ability to measure reflection coefficients of components is essential for absolute calibration of the LEDA radiometers. To facilitate this, Hirose MS147 test ports have been added to the circuit. When a MS147 cable is connected to the test port, an internal mechanical switch within the test port reroutes the circuit to the MS147 cable. The MS147 ports are used to measure the reflection coefficients, $\ss{\Gamma}{ant}$ and $\ss{\Gamma}{rx}$, as introduced in Sec.~\ref{sec:calibration}.

\subsubsection{Calibration subcircuit} \label{sub:cal}

%%% FIGURE
\begin{figure*}
  \centering
  \includegraphics[width=1.6\columnwidth]{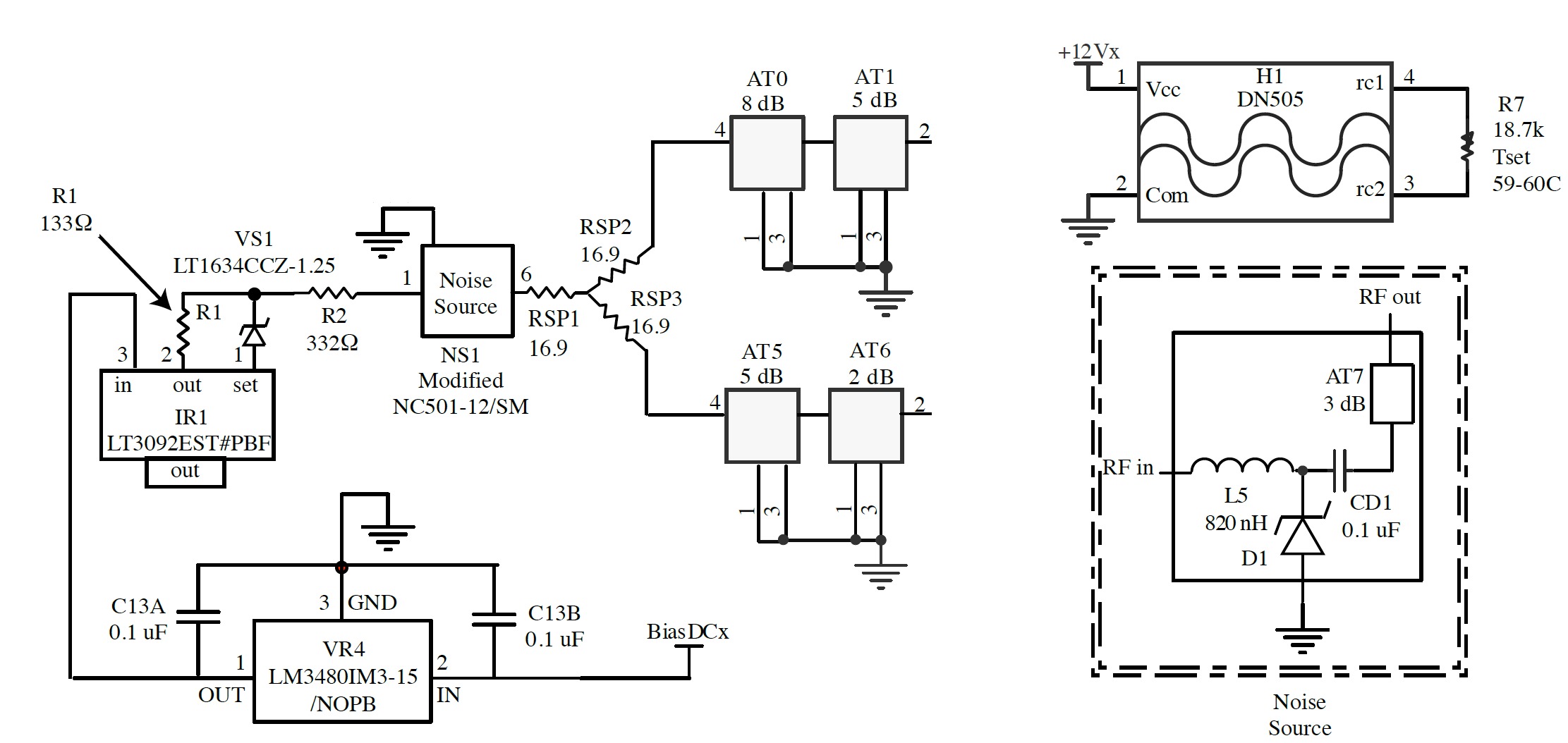}
  \caption{Circuit schematic for the calibration subcircuit. This circuit is housed in a Delrin shell; the DN505 heater is placed on top of the modified NC501 package. \label{fig:diode}}
\end{figure*}

At LEDA frequencies, the intrinsic sky noise  (>1000\,K) is much greater than ambient temperature ($\sim$300 K). As such, comparison to an ambient 50\,$\Omega$ load results in a large swing in LNA input power; this is undesirable for two reasons. Firstly, digitizer dynamic range requirements are reduced, allowing more overhead to deal with radio interference. Secondly, large swings in power change which bits in the digitizer are being exercised, and are consequently more likely to be affected by non-linear quantization gain. Additionally, the signal to noise ratio after applying three-state switching (Eq.~\ref{eq:residuals-hot-cold-1}) is improved by using stronger references. As such, the reference calibration states on the FE are provided by a noise diode based subcircuit, with equivalent noise temperatures better matched to the sky temperature.

The calibration subcircuit is shown in Fig.~\ref{fig:diode}. This circuit is based on a NoiseWave NC501-12/SM noise diode package, which we have modified to improve its stability. The top plastic cover of the NC501 was removed and replaced by a DN505 heater seated upon  an aluminium block. The DN505 is set to maintain a constant temperature of 60\,$^\circ$C, to mitigate ambient temperature variations. The circuit includes a constant voltage regulator (LM3480) and constant current regulator (LT3092) to ensure the noise diode receives a stable DC supply.

A resistive splitter is used to provide two calibration paths, upon which we place different attenuators to yield `hot' and `cold' references. Although a resistive splitter gives 3 dB of loss as compared to an ideal power divider, we found that thermal stability---the change in noise power output as a function of ambient temperature---was notably worse when a power divider was used; see Sec.~\ref{sub:diode-stability} for measurement details.

\subsection{Analog backend} \label{sec:analog}
%%% FIGURE
\begin{figure}
  \centering
  \includegraphics[width=1.0\columnwidth]{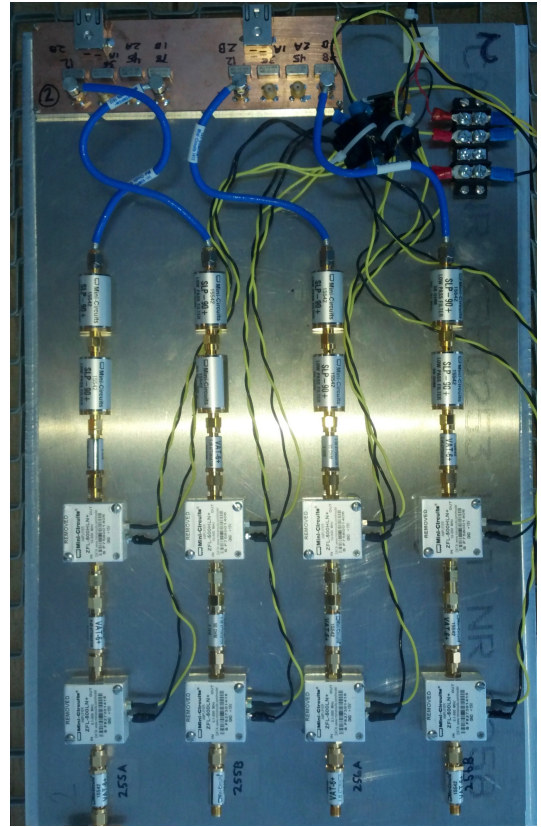}
  \caption{Four of ten signal paths in the LEDA connectorized backend electronics (CRX), which was adopted during an upgrade cycle in place of closely packed and unshielded LWA RF circuitry to reduce pickup of  internally generated RFI and reflections. \label{fig:crx}}
\end{figure}

Second-stage signal conditioning for the LEDA radiometers is performed using a connectorized analog receiver chain (CRX, Fig.~\ref{fig:crx}). While the station is outfitted with an LWA 512-input second-stage analog receiver system (ARX) \citep{ellingson2013}, used to amplify and filter antenna signals prior to correlation, this system was not suitable for  precision radiometry: the ARX demonstrated crosstalk between neighboring channels, pickup of radiation emitted by the densely packed signal paths inside the shielded analog rack, and reflection along the signal path that affected the noise floor.  

\subsubsection{Connectorized receiver system}

The first two components of the CRX are situated at the RF-shielded shelter's entry bulkhead: a Lightning arrestor (PolyPhaser GT-NFF-AL) and a FM bandstop filter (Mini-Circuits NBSP-108+). A RG316-DS coaxial cable connects the bulkhead components to the rest of the CRX, which is located in an RF-shielded rack. A bias tee (Mini-Circuits ZFBT-282-1.5A+) supplies DC voltage back to the antenna, while blocking DC on the output port. A highpass filter (Mini-Circuits SHP-50+) filters out VHF interference, this is connected to a Mini-Circuits ZFL-500LN+ amplifier ($G$=24\,dB). A further 19\,dB of amplification is provided by a ZFL-500HLN+ amplifier, which is optimized for higher power than the preceding 500LN+ model. Final-stage filtering of the FM band and above is then provided by two Mini-Circuits SLP-90+ filters in cascade. For improved impedance matching, Mini-Circuits VAT-6+ 6\,dB attenuators are placed before, between, and after the two amplifiers.

\subsubsection{Switching Assembly}

In order to select between sky and reference diode states on the FE, different DC power levels are supplied via the CRX bias tees. The DC power supplied to the CRX bias tees is controlled by the switching assembly (SAX), shown in Fig~\ref{fig:block_diag} as the receiver board voltage controller. The SAX system consists of a custom voltage regulation circuit controlled by a Rabbit 3000 Microprocessor via a wired Ethernet connection. The SAX accepts DC input power from a 28\,V, 11\,A Acopian Gold linear supply (A28H1100-230), and outputs power at 17, 20 and 23\,V, as required by the FE. Tunable potentiometers may be used to adjust the output power, to account for power drop over the coaxial cables (1--1.5\,V).

The SAX accepts a PPS signal, which may be used to trigger state changes on the FE. Alternatively, FE state may be controlled manually, by issuing commands to the Rabbit microprocessor over Ethernet. As the SAX is located in the analog rack, it is encased in an RF-tight box, to shield any microprocessor-generated RF power from the analog systems.

\subsection{Inter-rack signal transport}

The digital and analog systems are housed in separate RF-shielded racks that share a common bulkhead wall.  Consistent with the LWA engineering model, CRX signals in the analog rack are converted from unbalanced 50\,$\Omega$  to balanced 100\,$\Omega$ for transmission to the digital rack on Category-7a (CAT7A) Ethernet cable.  A custom-made balun module converts every 4$\times$SMA inputs into a single Belfuse ARJ45-ended output (one  conductor pair per RF signal path).  Custom RF-tight, CAT-7A CONEC feedthroughs route signals through bulkhead plates.

ARJ45 was substituted for LWA-standard RJ45 CAT5e during an upgrade cycle, motivated by superior near-end crosstalk (NEXT).  Although the digitizer cards were not upgraded, VNA S21 bench measurements demonstrated cross-talk of $< -60$ dB  for a CAT7A signal path including a passthrough and single CAT5e plug and jack combination at one end.  With this, residual cross talk in the signal path is dominated by the ADC card. Bench measurements have demonstrated reduction to between $-30$ and $-40$ dB across the science band when only two of four conductor pairs are used on each cable, corresponding to CAT5e 8P8C connector pins 1, 2 and 7, 8 which are physically the most widely separated.  

\section{Calibration equations} \label{sec:calibration}

%%%% FIGURE

\begin{figure}
  \centering
  \includegraphics[width=1.0\columnwidth]{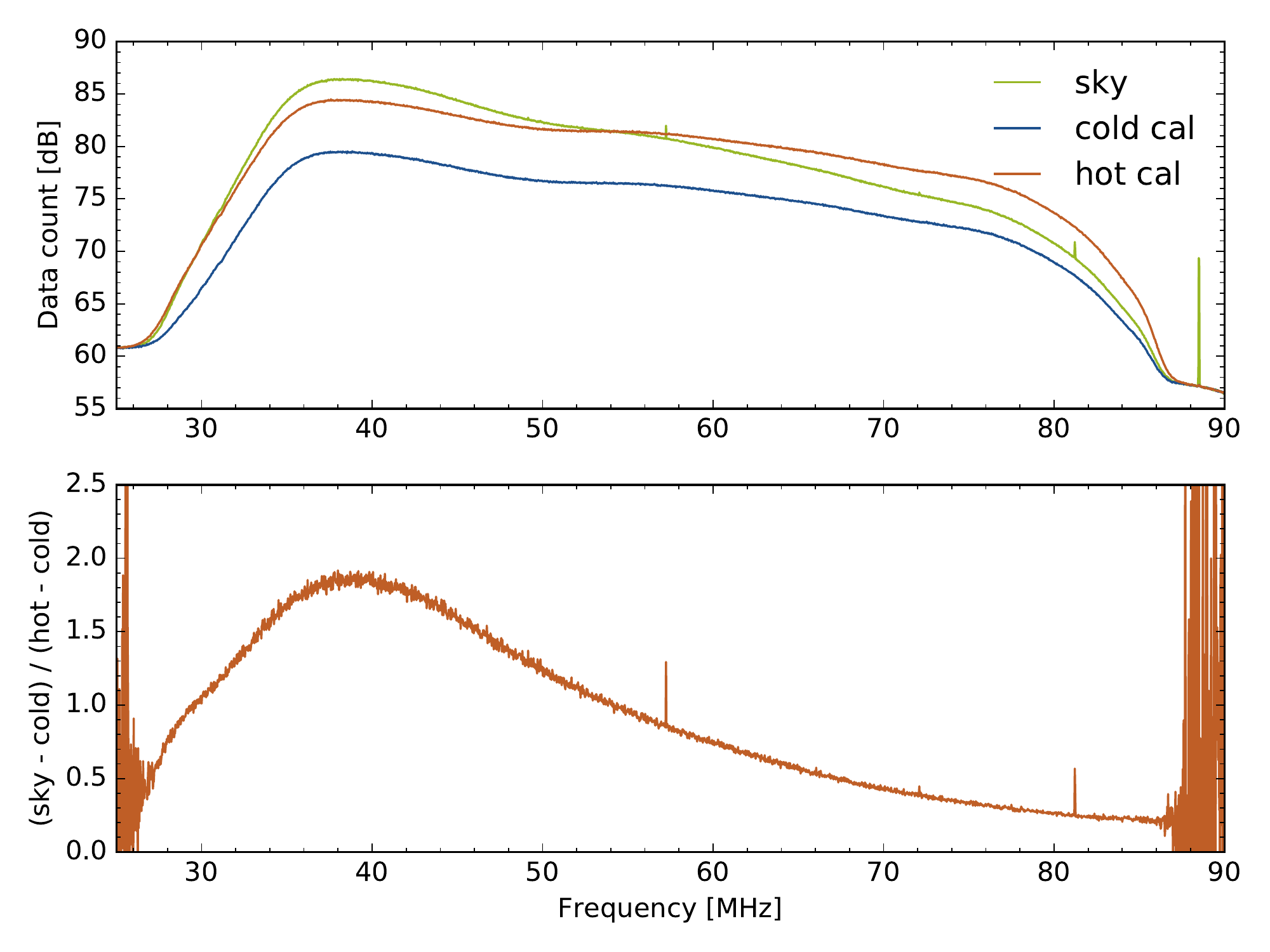}
  \caption{Top: typical bandpasses of antenna 252A in sky and calibration states, i.e. $P_{\rm{ant}}$, $P_{\rm{cold}}$ and $P_{\rm{hot}}$. 
Bottom: the ratio $(P_{\rm{ant}} - P_{\rm{cold}}) / (P_{\rm{hot}} - P_{\rm{cold}})$ of data in top panel.  \label{fig:shc}}
\end{figure}

The output power of a radiometer system is given by
\begin{equation}
P_{\rm{out}} = G k_B \Delta\nu (T_{\rm{ant}} + T_{\rm{rx}}) \label{eq:radpow}
\end{equation}
where $G$ is the total gain of the radiometer's analog systems (amplifiers, filters, etc), $k_B$ is Boltzmann's constant, and $T_{\rm{rx}}$ and $T_{\rm{ant}}$ are the noise-equivalent temperatures of the receiver and antenna, respectively.  At frequencies corresponding to the CD trough, radiometer systems are generally sky-noise dominated; that is, $T_{\rm{ant}} \gg T_{\rm{rx}}$.  Eq.~\ref{eq:radpow} is the fundamental measurement made by a radiometer.

Antenna temperature is given by the average of the actual sky brightness $T_{sky}(\theta,\phi)$
as seen from the antenna's location, weighted by the antenna's gain pattern
$B(\theta,\phi)$:
\begin{equation}
T_{\rm{ant}}(\nu)=\frac{\int d\Omega B(\theta,\phi,\nu)T_{\rm{sky}}(\theta,\phi,\nu)}{\int d\Omega B(\theta,\phi,\nu)}.\label{eq:sky}
\end{equation}
If one separates the sky temperature $T_{\rm{sky}}$ into a "foreground" component, $T_{\rm{fg}}$, and the cosmological, $T_{\rm{cosm}}$, term consisting of the sky-averaged 21-cm emission from Eq. \ref{Eq:T} and the background CMB radiation, then
\begin{equation}
T_{\rm{sky}}(\theta, \phi, \nu) = T_{\rm{fg}}(\theta, \phi, \nu) + T_{\rm{cosm}}(\nu),
\label{eq:Tsky}
\end{equation}
such that Eq.~\ref{eq:sky} becomes
\begin{equation}
T_{\rm{ant}}(\nu)=\frac{\int d\Omega B(\theta,\phi,\nu)T_{\rm{fg}}(\theta,\phi,\nu)}{\int d\Omega B(\theta,\phi,\nu)} + T_{\rm{cosm}}(\nu).\label{eq:fgbg}
\end{equation}
In practice, impedance mismatch between the antenna and receiver must also be taken into consideration, as must noise waves generated by the receiver's first-stage amplifier. These factors are detailed further in Sec.~\ref{sec:calibration}.

The LEDA radiometer employs a three-state switching calibration technique \citep{rogers2012}, where the receiver cycles between the sky and two calibration references (a "hot" and "cold" state). Three-state switching allows the removal of variations in system gain $G = G(\nu, t)$ and receiver temperature $\ss{T}{rx} = \ss{T}{rx}(\nu,t)$, and allows for a temperature scale to be imposed on the data. 

The LEDA outrigger antennas switch between the sky and two calibration reference paths. A noise diode in series with attenuators is used to provide a reference with an equivalent noise temperature of $\ss{T}{hot}$ and $\ss{T}{cold}$. The power measured by the radiometer in each state (Eq.~\ref{eq:radpow}) is given by
\begin{align}
P_{\rm{ant}} & = Gk_{\rm{B}}\Delta\nu(\ss{T}{ant}+\ss{T}{rx})\\
P_{\rm{hot}} & = Gk_{\rm{B}}\Delta\nu(\ss{T}{hot}+\ss{T}{rx})\\
P_{\rm{cold}}& = Gk_{\rm{B}}\Delta\nu(\ss{T}{cold}+\ss{T}{rx}),
\end{align}
where $P_{\rm{ant}}$,  $P_{\rm{hot}}$ and $P_{\rm{cold}}$ are powers measured in antenna, hot reference, and cold reference states;  $T_{\rm{ant}}$,  $T_{\rm{hot}}$ and $T_{\rm{cold}}$ are antenna, hot and cold reference noise-equivalent temperatures. The three-state switch calibrated temperature $T_{\rm{ant}}$ may then be recovered via
\begin{equation}
\ss{T}{ant} = (\ss{T}{hot} - \ss{T}{cold})\frac{P_{\rm{ant}}-P_{\rm{cold}} }{P_{\rm{hot}}-P_{\rm{cold}}}+T_{\rm{cold}}.\label{eq:3ss}
\end{equation}
Example spectra for the three states, is shown in Fig.~\ref{fig:shc}. 
As presented in \cite{rogers2012}, the true antenna temperature  $T_{\rm{cant}}$  is related to the three-state calibrated  $T_{\rm{ant}}$ by 
\begin{equation}
T_{\rm{cant}}=T_{\rm{ant}}(1-|\Gamma|^{2}),\label{eq:vna}
\end{equation}
where  $\Gamma$ is the reflection coefficient: a measure of impedance mismatch between the receiver and the antenna. 

However, Eq.~\ref{eq:vna} is not strictly accurate for two reasons. Firstly, one must take care to use an appropriate definition for power gain $G$, for which there are several (see e.g., \cite{pozar2005}). Here, we are interested in the power delivered to the load from a given source, for which the \emph{transducer power gain} should be used. As shown in \cite{pozar2005}, for a given amplifier with $S_{12}$ (reverse isolation, see Section \ref{sec:scat} for more details) negligibly small, when connected to a source with reflection coefficient $\Gamma_{S}$ and a load with reflection coefficient $\Gamma_{L}$, the transducer power gain is given by:
\begin{equation}
G_{T}=\frac{|S_{21}|^{2}(1-|\Gamma_{S}|^{2})(1-|\Gamma_{L}|^{2})}{|1-S_{11}\Gamma_{S}|^{2}|1-S_{22}\Gamma_{L}|^{2}},
\label{eq:gain-transducer}
\end{equation}
where  $S_{21}$  is a parameter equivalent to forward gain, and $S_{11}$ and $S_{22}$ are reflection coefficients.  
Note that in the ideal case, $\Gamma_L=\Gamma_S=0$ and Eq.~\ref{eq:gain-transducer} yields $G_T=|S_{21}|^2$.

Secondly, the noise temperature of the receiver, $T_{{\rm rx}}$, also depends upon the source. For an amplifier with optimal noise figure $F_{{\rm opt}}$, the noise figure F for a given $\Gamma_{S}$ is given by
\begin{equation}
	F=F_{{\rm opt}}+\frac{4R_{N}}{Z_{0}}\frac{|\Gamma_{S}-\Gamma_{{\rm opt}}|^{2}}{(1-|\Gamma_{S}|^{2})|1+\Gamma_{opt}|^{2}} \label{eq:noise-gamma}
\end{equation}
where $R_{N}$ is the equivalent noise resistance of the amplifier, $Z_{0}$ is the characteristic impedance, and $\Gamma_{opt}$ is the amplifier's reflection coefficient at which its noise figure is the lowest. The receiver temperature---defined as $T_{rx}=T_{0}(F-1)$ with $T_{0}$=290\,K---is thus dependent upon the source. An alternative approach to modeling noise within an analog system is to use noise correlation matrices \citep{king2010, king2014}, or `noise wave' analysis \citep{meys1978}; the EDGES formalism uses the latter approach.

The magnitude of the inaccuracy of Eq.~\ref{eq:3ss} due to neglect of Eq.~\ref{eq:gain-transducer} and Eq.~\ref{eq:noise-gamma} is primarily dependent upon $\Gamma_S$. The formalism of \cite{rogers2012} (and \cite{Monsalve:2017a}) is therefore only accurate in the case that the diode and load states are well matched to the receiver; in both the LEDA and EDGES instruments, $\Gamma_S$ is lower than $-30$\,dB for reference states, so this requirement is satisfied. 

For the calibration detailed here, we follow the formalism of \cite{rogers2012} and \cite{Monsalve2017a}, but nonetheless highlight that improvement of the formalism is an area deserving future examination. Following this formalism, $\Gamma$ in Eq.~\ref{eq:vna}, is calculated from reflection coefficients of the antenna and receiver, $\Gamma_{\rm{ant}}$  and $\Gamma_{\rm{rx}}$ respectively (Fig.~\ref{fig:block_diag}), measured with a vector network analyzer (VNA), and
\begin{align}
\ss{T}{cant} & = \ss{T}{sky}\ss{H}{ant}|F|^{2}\ss{H}{rx}^{-1}\nonumber \\
 & + \ss{T}{u}|\ss{\Gamma}{ant}|^{2}|F|^{2}\ss{H}{rx}^{-1}\nonumber \\
 & +(\ss{T}{c} cos(\psi) + \ss{T}{s} sin(\psi))|\ss{\Gamma}{ant}||F|\ss{H}{rx}^{-1}.\label{eq:T_3ss}
\end{align}
The terms of Eq.~\ref{eq:T_3ss} are:
\begin{itemize}
\item $\ss{\Gamma}{ant}$ is the reflection coefficient of the antenna, as measured
at the output of the balun.
\item $\ss{\Gamma}{rx}$ is the reflection coefficient of the receiver. The first component in the receiver is a
low noise amplifier (LNA), which will be a main cause of reflections
between the antenna and the receiver.
\item $\ss{H}{ant}=1-|\ss{\Gamma}{ant}|^{2}$ and $\ss{H}{rx}=1-|\ss{\Gamma}{rx}|^{2}$ are gain terms arising due to antenna / receiver mismatch.
\item $F=(1-|\ss{\Gamma}{rx}|^{2})^{1/2}(1-\ss{\Gamma}{ant}\ss{\Gamma}{rx})^{-1}$ is another
complex gain factor encompassing receiver and antenna mismatch.
\item $T_{u}$ is the uncorrelated `noise wave' power, emitted from
the LNA and reflected back by the antenna \citep[see][for noise wave formulation]{meys1978}.

\item $\ss{T}{c} \cos(\psi)$ and $\ss{T}{s} \sin(\psi)$ are correlated noise waves
that depend on the amplitude and phase of the antenna reflection, where 
$\psi$ is the phase of the noise wave reflected from the antenna.
\end{itemize}
From Eq.~\ref{eq:T_3ss}, one may solve for the true sky temperature $\ss{T}{sky}$. If the antenna is not lossless, a further correction must be applied:
\begin{equation}
\ss{T}{csky} =(\ss{T}{sky} - \ss{T}{amb}(1-L))/L,
\end{equation}
where $L=10^{-l/10}$ for a loss $l$ in dB, and $T_{amb}$ is the
ambient temperature of the antenna.

\subsection{Thermal uncertainties}

The LEDA receiver uses two reference diode states, in contrast to the load and diode approach used in EDGES. Here we show that the dual diode approach optimizes measurement signal to noise. 

\begin{figure}
  \centering
  \includegraphics[width=1.0\columnwidth]{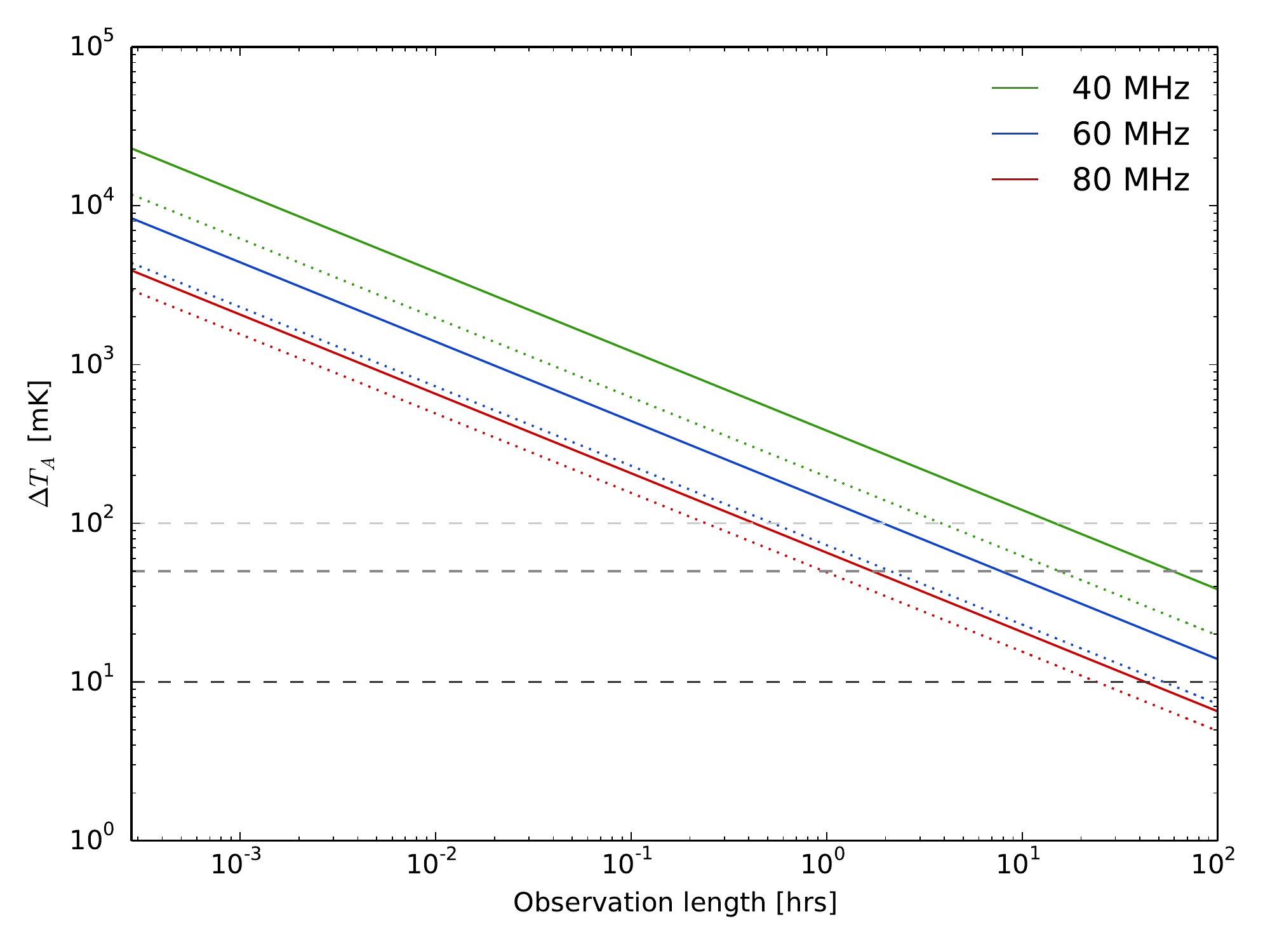}
  \caption{Expected improvement of measurement residuals for a system using dual noise diode references, based on use of Eq.~\ref{eq:rms-unc}. The dotted lines represent a system with $T_{\rm{hot}}=6500$\,K and $T_{\rm{cold}}=1000$\,K; solid lines represent a system with $T_{\rm{hot}}=450$\,K and $T_{\rm{cold}}=300$\,K. \label{fig:diode-rms}}
\end{figure}

An estimate of thermal noise present in the three-state switched spectrum may be found by propagating the uncertainties of Eq.~\ref{eq:3ss}:
\begin{equation}
dT_{\rm{ant}}^{2}=\left(\frac{\partial T_{\rm{ant}}}{\partial P_{\rm{ant}}}\right)^{2}dP_{\rm{ant}}^{2}+\left(\frac{\partial T_{\rm{ant}}}{\partial P_{\rm{cold}}}\right)^{2}dP_{\rm{cold}}^{2}+\left(\frac{\partial T_{\rm{ant}}}{\partial P_{\rm{hot}}}\right)^{2}dP_{\rm{hot}}^{2}.
\label{eq:rms-unc}
\end{equation}
where the uncertainty of measurement in each state is given by the radiometer equation. This yields
\begin{align}
dT_{{\rm ant}} & = A \sqrt{dP_{{\rm ant}}^{2}+B(dP_{{\rm hot}})^{2}+C(dP_{{\rm cold}})^{2}}\label{eq:residuals-hot-cold-1}\\
A & =\frac{T_{{\rm hot}}-T_{{\rm cold}}}{P_{{\rm hot}}-P_{{\rm cold}}} \\
B & =\left(\frac{P_{{\rm ant}}-P_{{\rm cold}}}{P_{{\rm hot}}-P_{{\rm cold}}}\right)^{2}\\
C & =\left(\frac{P_{{\rm ant}}-P_{{\rm hot}}}{P_{{\rm hot}}-P_{{\rm cold}}}\right)^{2}
\end{align}
It follows that to optimize measurement signal to noise, the two references should be as high power as possible, while maintaining a large difference in power between them. In tension with this, the finite dynamic range of the ADC motivates diode temperatures comparable to the sky brightness. Fig.~\ref{fig:diode-rms} compares the measurement uncertainty as a function of observation time for a dual diode system with fiducial values $T_{\rm{hot}}$=6500\,K and $T_{\rm{cold}}$=1000\,K, against a system with $T_{\rm{hot}}$=450\,K and $T_{\rm{cold}}$=300\,K. 

\section{Instrument characterization} \label{sec:char}

Of particular importance to calibration is a sound understanding of the characteristics of the FE. In this section, we present detailed measurements of the characteristics of the FE and antenna for a single polarization (antenna 252A). Nonetheless, all FE boards undergo the same characterization process; comparison between antennas is presented in following sections.

\subsection{Gain linearity}

\begin{figure}
\includegraphics[width=1\columnwidth]{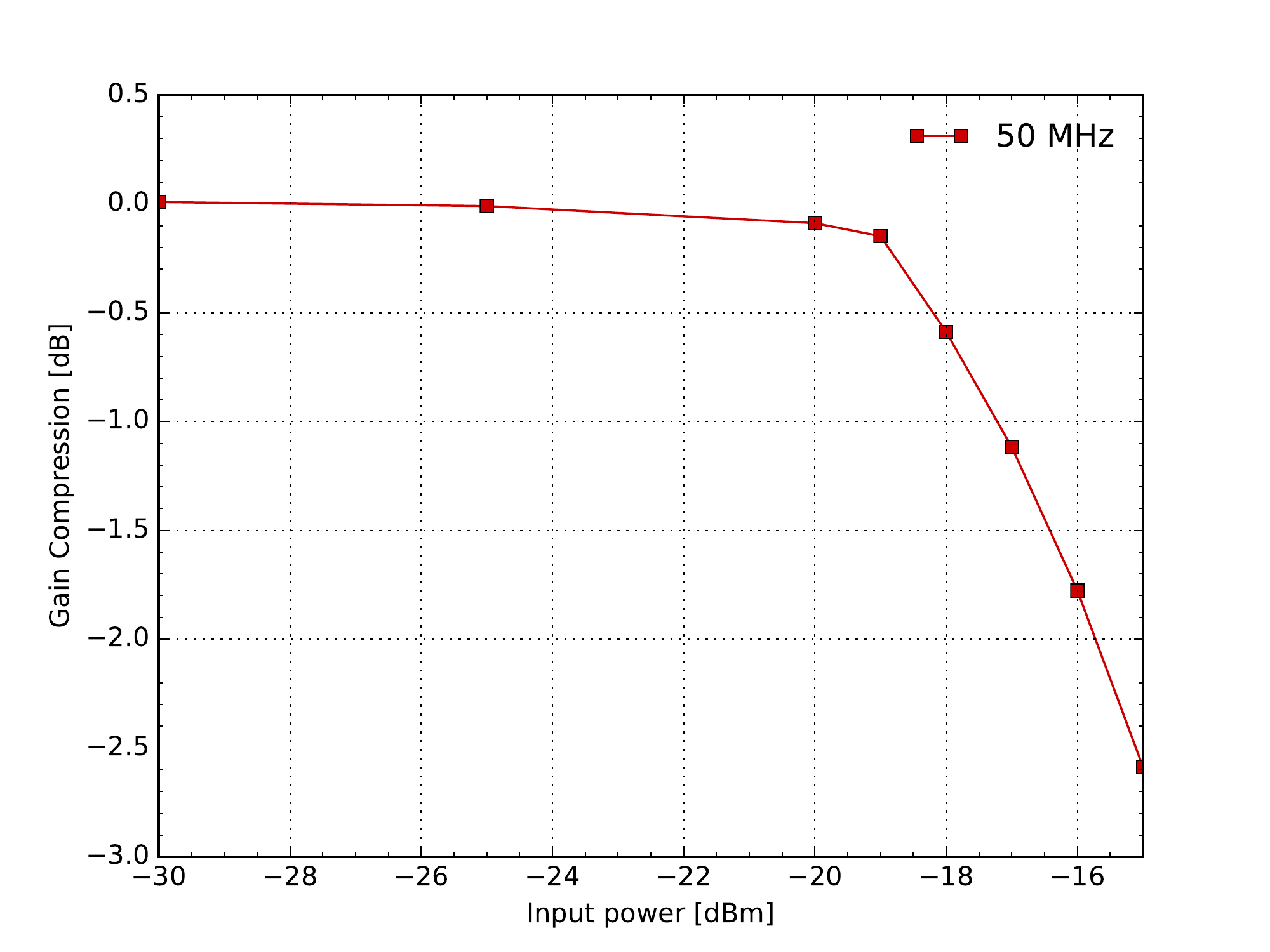}
\protect\caption{Gain compression curve for the LEDA receiver at 50 MHz.
1\,dB compression occurs at an input power of -17.2\,dBm. \label{fig:comp}}
\end{figure}

To ensure the amplifiers on the FE are operating in a nominal regime, we tested the gain linearity of the FE using an Agilent E4424B signal generator and Agilent N9000A spectrum analyzer. The signal generator was used to produce a 50\,MHz tone with amplitudes covering $-100$ to $-15$\,dBm. We found the 1\,dB compression point occurs at an input power of $-17.2$\,dBm (Fig.~\ref{fig:comp}). Note that the expected power from the antenna is well under this 1\,dB compression point: for a 10,000\,K sky over 100\,MHz (an overestimation), one would expect $-78.6$\,dBm input power to the FE. 

The first harmonic of the 50\,MHz tone was not apparent on the spectrum analyzer until an input power of $-40$\,dBm (far above expected input power), when it appeared above the spectrometer's noise floor with an output power of $-72.8$\,dBm. From our data, we extrapolate the IP2 intercept to be $\sim$37\,dBm.

%%%% FIGURE

\subsection{Scattering parameters}
\label{sec:scat}
The reflection and transmission characteristics of the LEDA FE and LWA antenna were measured using an Anritsu MS2034B Vector Network Analyzer (VNA). Between the MS147 test ports and the SMA output, the LEDA FE can be treated as a 2-port network. This allows us to measure its scattering parameters (S-parameters) relating incident and reflected voltage waves.

\subsubsection{Antenna + balun}
%%%% FIGURE
\begin{figure}
\includegraphics[width=1\columnwidth]{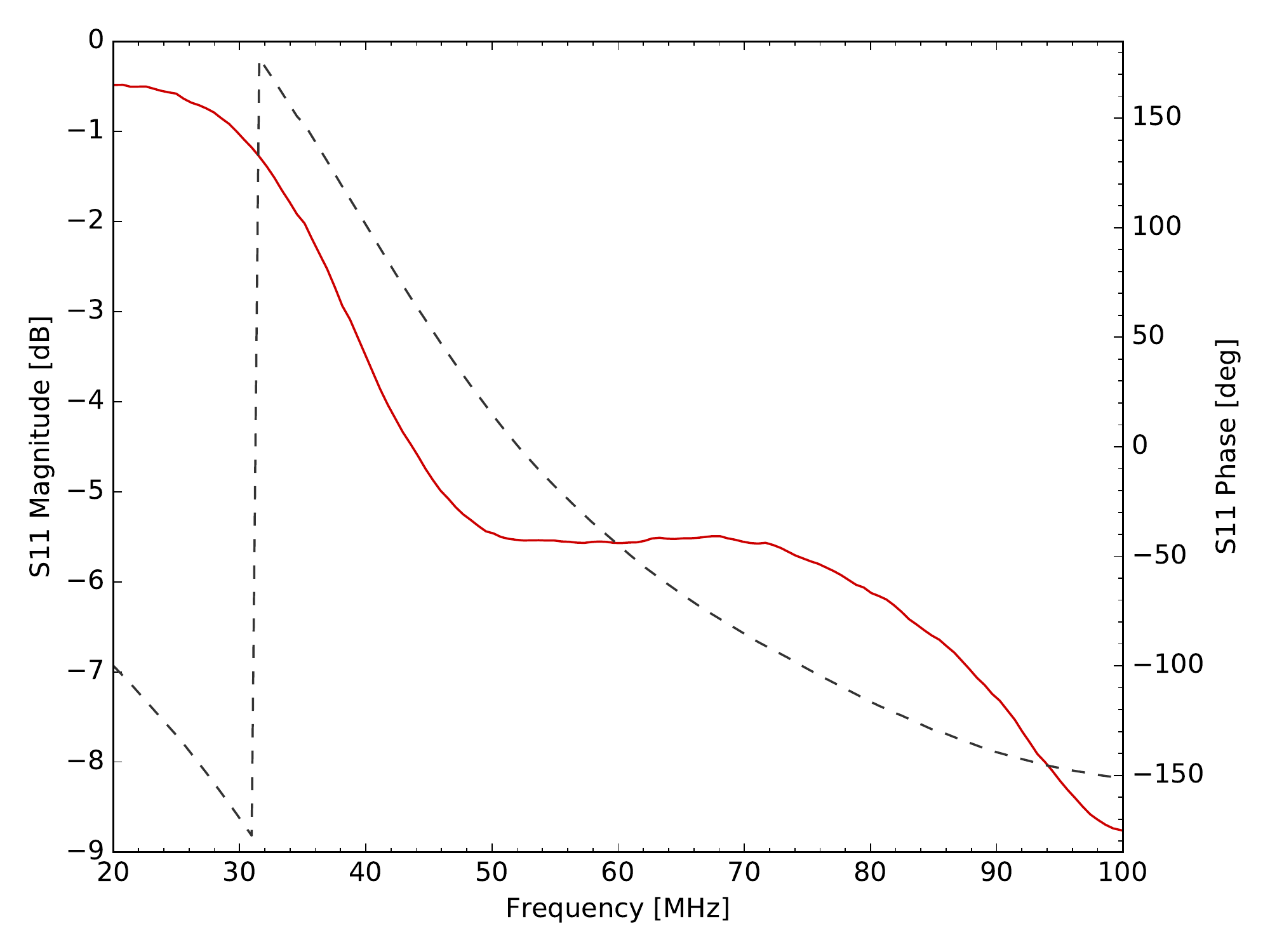}
\protect\caption{Magnitude (red) and phase (black dashed) of reflection coefficient $\Gamma_{\rm{ant}}$ for antenna 252A. These data were taken in January 2016. \label{fig:ants11}}
\end{figure}

A VNA measurement of the antenna cannot be made without the use of a balun; the characteristics of the balun must therefore be known for one to de-embed its effect. The MS147 connector directly after the FE balun allows the reflection coefficient $\ss{\Gamma}{ant}$ to be measured, see Fig.\,\ref{fig:block_diag}. 

Given the far-field distance of the antennas is many hundreds of meters, and that the surrounding environment affects the beam characteristics, data were necessarily taken {\emph{in-situ}} at OVRO. Care was taken to ensure that during measurement, all equipment was placed low to the ground as far away as possible. The VNA was placed on the ground, at a distance of 20\,m away from the antenna, orthogonal to the antenna blade pair under test. Low-loss coaxial cable was laid from the VNA, across the ground, and up the antenna's central mast to the FE MS147 connector that connects to the balun.  

The magnitude and phase of $\ss{\Gamma}{ant}$ for antenna 252A is shown in Fig.~\ref{fig:ants11}. The magnitude and phase of $\ss{\Gamma}{ant}$ are seen to vary smoothly as a function of frequency, varying between $-4$ to $-6$\,dB over the 40--85\,MHz band.

\subsubsection{Front-end receiver}
%%%% FIGURE
\begin{figure*}
\includegraphics[width=1.5\columnwidth]{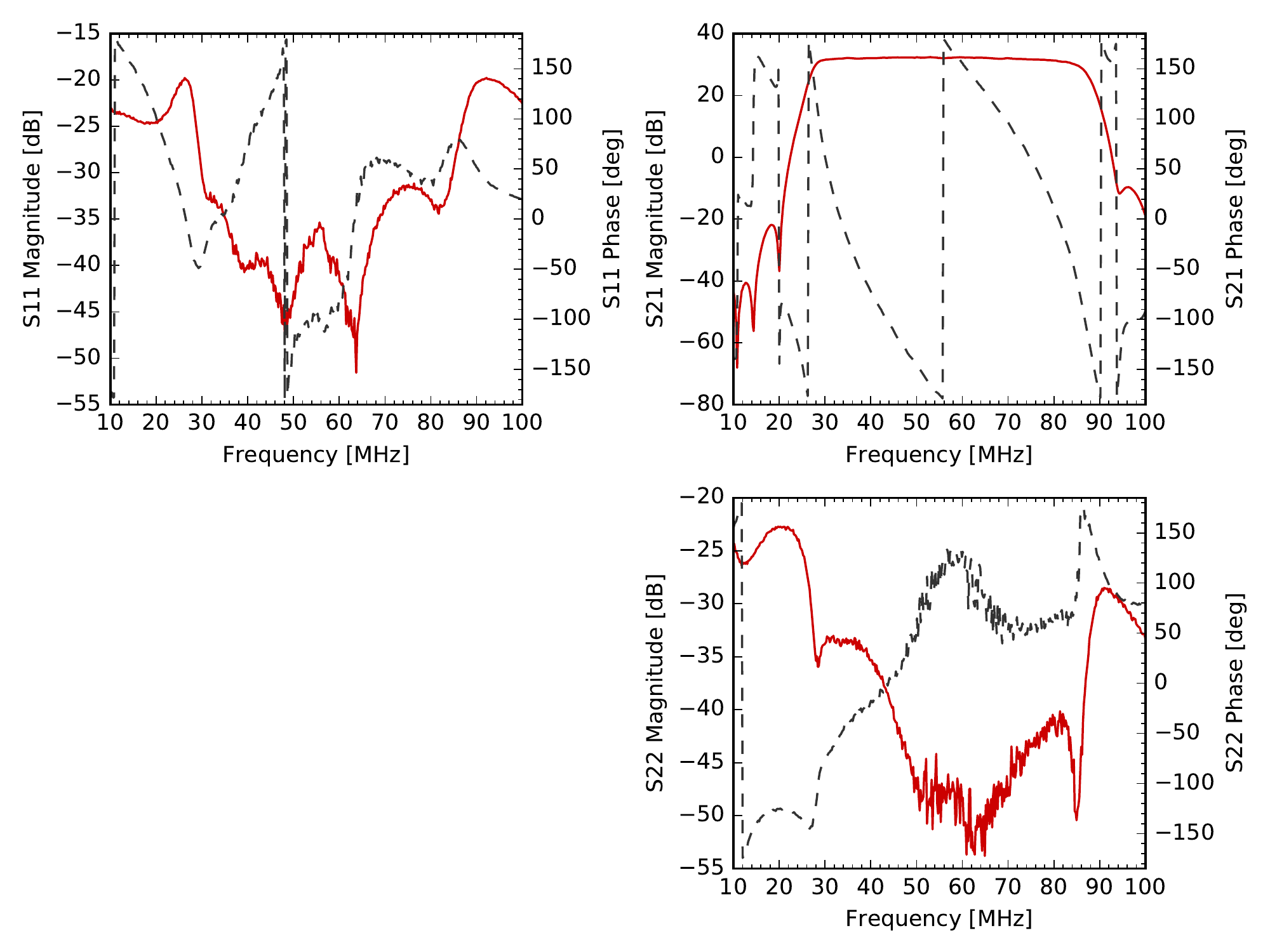}
\protect\caption{Magnitude (red) and phase (dashed black) of scattering parameters for LEDA FE installed on antenna 252A. The $S_{\rm 12}$ parameter is below the measurement ability of the VNA (below $-60$\,dB) so is not shown.\label{fig:fes11}}
\end{figure*}

S-parameter measurements of the FE were taken by connecting the VNA to the MS147 connector directly preceding the switch. This port is labelled  $\ss{\Gamma}{sw0}$ in Fig.~\ref{fig:block_diag}; we will refer to this as $\ss{\Gamma}{rx}$, as it is the main VNA measurement presented for the receiver board. The board was characterized over 10--100\,MHz, using a low VNA port power ($-25$\,dBm) such that the FE was operating in a linear gain regime. So that the FE could be measured as a single device-under-test (DUT), the board was powered via the regulator daughter board in lieu of using an external bias tee. 

The S-parameters for the FE are shown in Fig.~\ref{fig:fes11}. The overall gain ($S_{\rm 21}$) is $32.1\pm0.2$\,dB over 30--80\,MHz. The 3\,dB and 10\,dB rolloff points occur at (27.7, 85.5)\,MHz and (26.0, 88.7)\,MHz, respectively; primarily due to the bandpass filter. The $S_{11}$  (i.e. $\ss{\Gamma}{ant}$), is better than $-30$\,dB across the LEDA science band of 40--85\,MHz, with $S_{11}$ increasing outside of the passband. Similarly, the $S_{22}$ is also better than $-30$\,dB across the LEDA science band.

\subsubsection{Calibration parameters}

\begin{figure*}
\centering
\includegraphics[width=2.0\columnwidth]{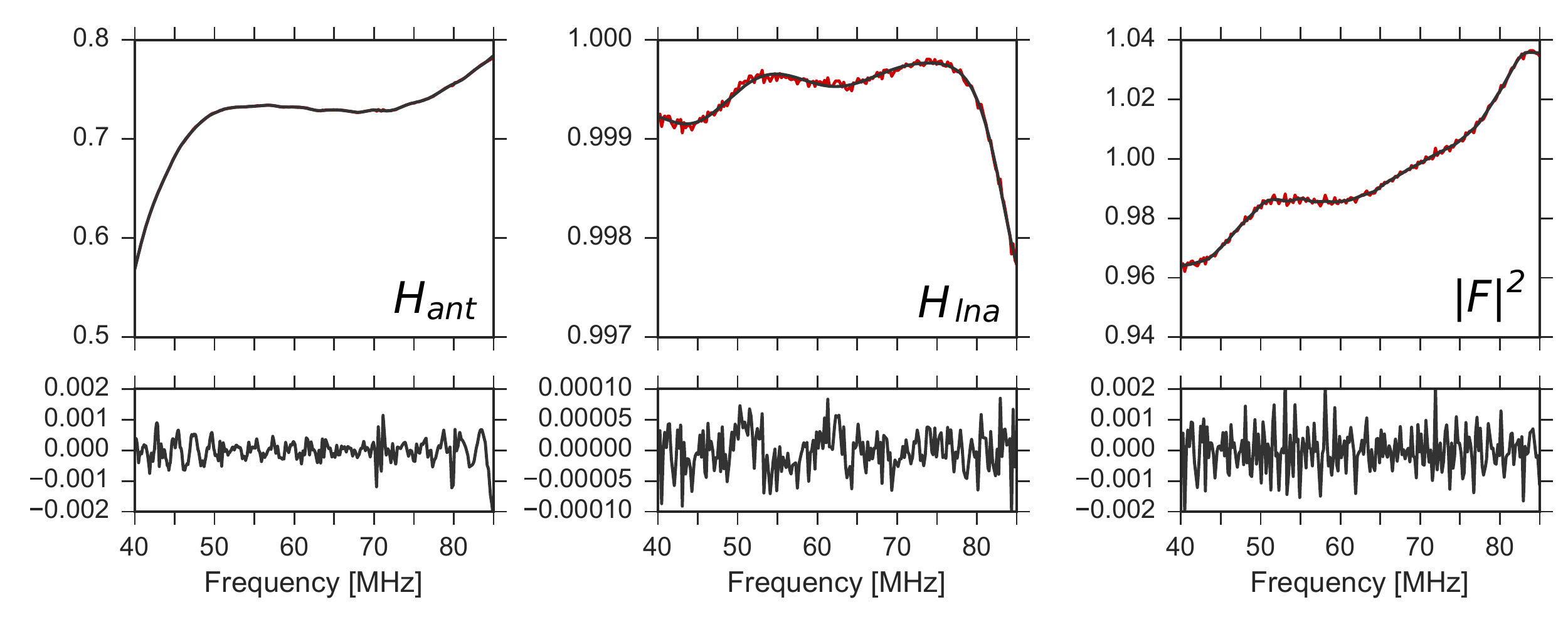}
\protect\caption{ 
	Calibration parameters $\ss{H}{ant}$, $\ss{H}{rx}$, and  $|F|^2$ (left to right). In the top panels, VNA measurements are shown in red, with a model fitted in black. The bottom panel shows the residual between the VNA measurements and the model.
	\label{fig:cal-params}
}
\end{figure*}

From the VNA measurements of the FE and antenna, we are able to form the calibration parameters $\ss{H}{ant}$, $\ss{H}{rx}$ and $|F|^2$ (Fig.~\ref{fig:cal-params}). In the top panels of Fig.~\ref{fig:cal-params}, the VNA measurements are shown in red, and smoothly-varying fitted models are shown in black; the bottom panel shows the residual between the VNA data and the fit. We have fit $\ss{H}{rx}$ with an 11-term polynomial, $|F|^2$ with a 21-term Fourier series, and $\ss{H}{ant}$ with a combined 5-term polynomial and 21-term Fourier series. As such, $\ss{H}{ant}$ has the largest effect upon calibration; the LNA's low reflection coefficient $\ss{\Gamma}{rx}$ means that $\ss{H}{rx}$ has only a small (<0.1\%) effect upon the overall calibration.

\subsection{Receiver temperature}

We determined the receiver temperature $\ss{T}{rx}$ using the Y-factor method \citep{pozar2005}. 
For an accurate measurement, noise contribution and any loss from the cables and connectors between the reference source and DUT must be included in $\ss{T}{hot}$ and $\ss{T}{cold}$. Further, precise measurement requires $Y \gg 1$, meaning the hot and cold references should be as different as possible.

To measure the receiver temperature of the LEDA receiver boards (DUT), we applied the Y-factor method using a calibrated HP 346C noise source as a reference. Hot and cold references states were created by inserting a 10\,dB and 6\,dB pad between the 346C and the DUT; the resulting noise temperature was computed using
\begin{equation}
	T_{{\rm cal}}=(1-L_{att})T_{346C} + 290 L_{att}
\end{equation}
where $L_{\rm{att}}$ is the combined loss of the attenuator and coaxial cable, as measured using a VNA, and $T_{\rm{346C}}$ is the manufacturer specified noise temperature of the HP 346C source.

\begin{figure}
\includegraphics[width=1\columnwidth]{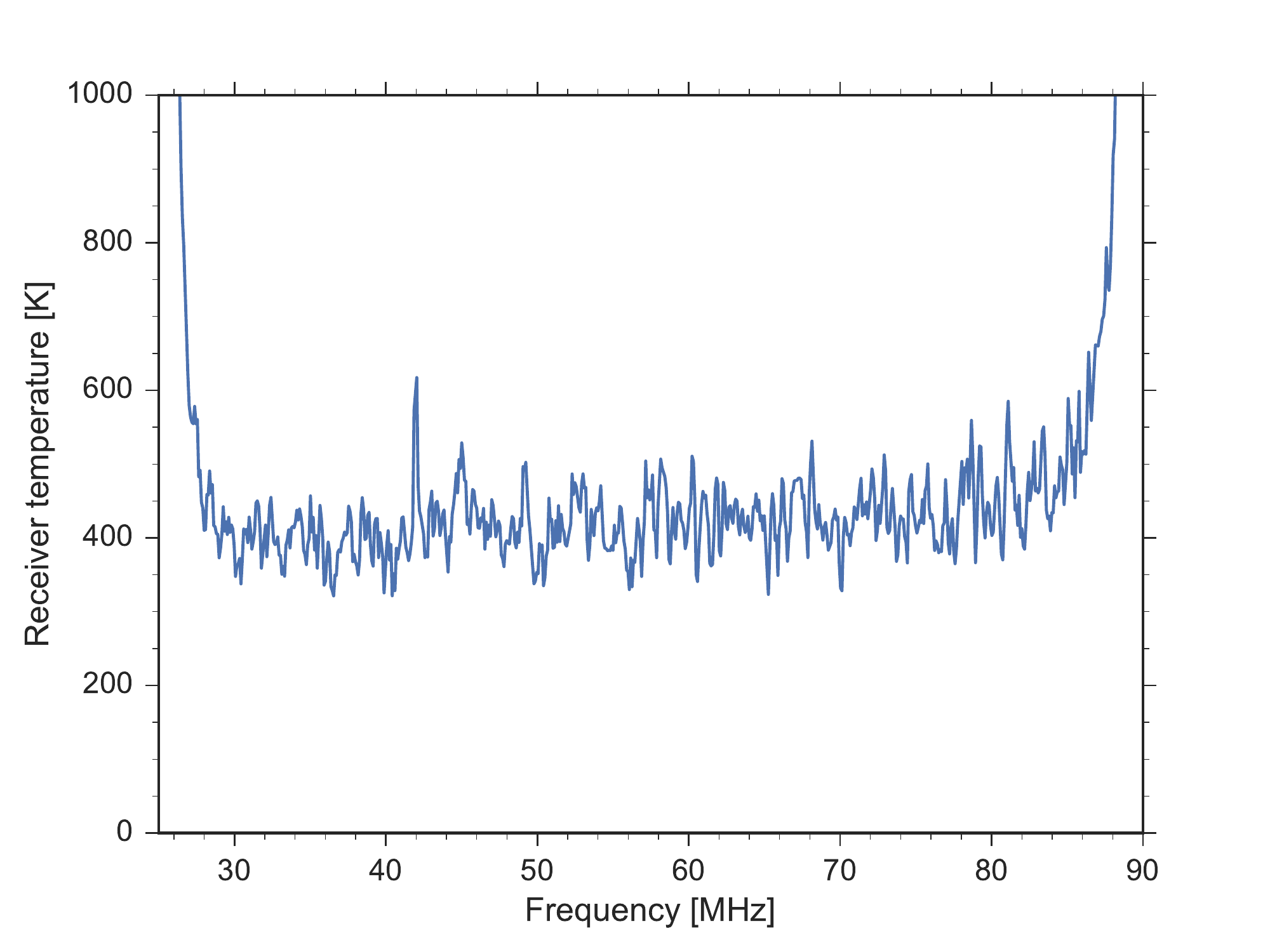}
\protect\caption{LEDA FE receiver temperature for antenna 252A. \label{fig:fee-rx}}
\end{figure}

We measured the $\ss{T}{rx}$ of all FE boards in the laboratory immediately prior to installation, using an Agilent 9000A spectrum analyzer; the HP346C reference source was connected at the MS147 test port between the balun and the switch. We find the receiver temperatures to be in good agreement with Tab.~\ref{tab:tsys} within 30--80\,MHz, increasing rapidly outside the receiver passband (\ref{fig:fee-rx}). For data analysis, we fit a line to the measured $\ss{T}{rx}$ between 40 and 80 MHz. 
 
%%%% FIGURE

\subsubsection{Noise diode temperatures}

\begin{figure}
\includegraphics[width=1\columnwidth]{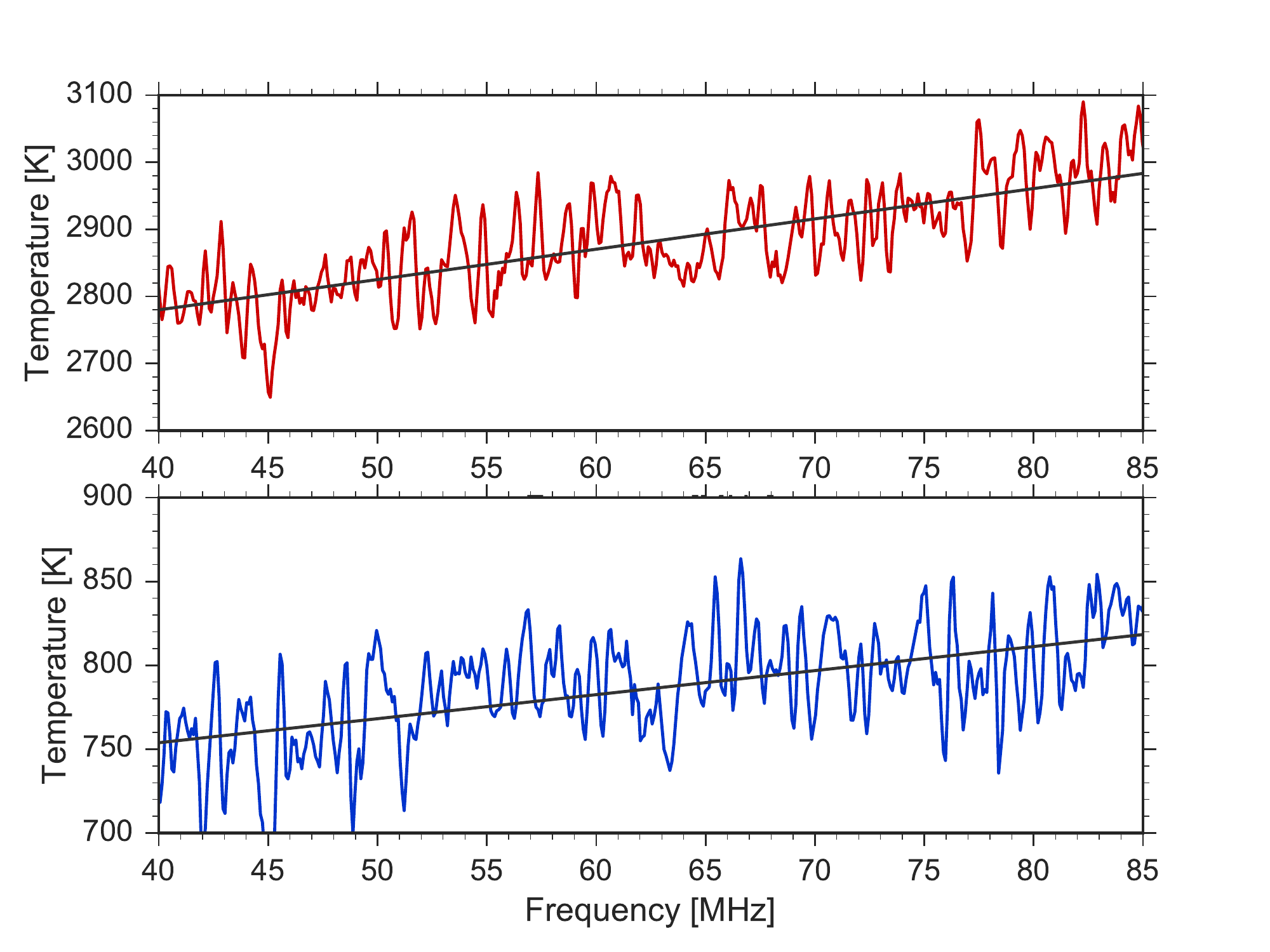}
\protect\caption{LEDA FE noise diode temperatures for antenna 252A. \label{fig:fee-diodetemp}}
\end{figure}

Once the receiver temperature is known, the temperature of the hot and cold references can be calculated with reference to the external HP346C. Fig.~\ref{fig:fee-diodetemp} shows the equivalent noise temperature of the hot (red) and cold (blue) noise diode reference states. A linear model is fitted to both states (black), which is used in subsequent calibration.

%%%% FIGURE
 
\subsection{LNA noise wave analysis}

\begin{figure}
\includegraphics[width=1\columnwidth]{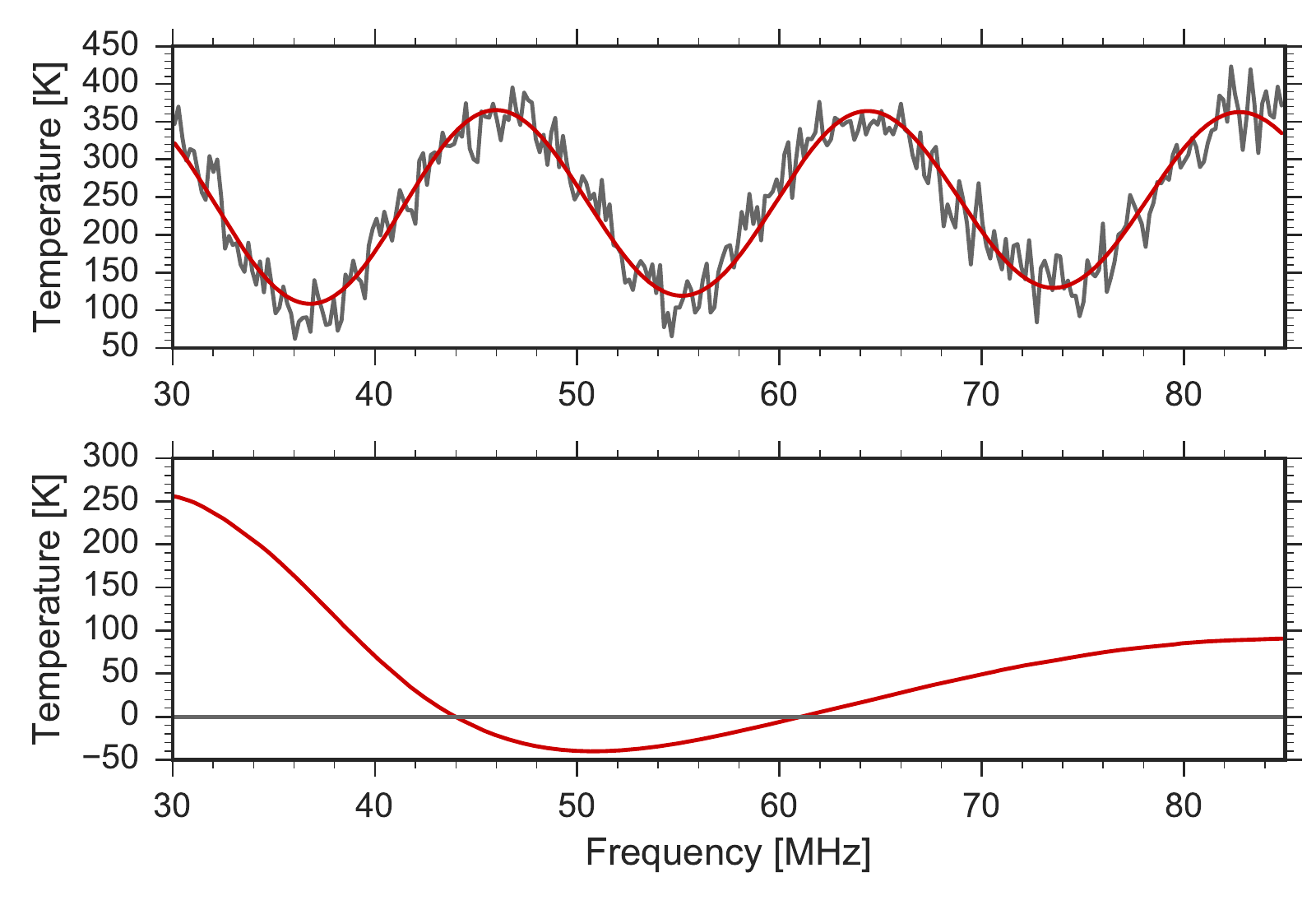}
\protect\caption{Top: spectra of coaxial cable used to characterize FE receiver noise wave, with measured spectra in grey and fitted model for scalar values of $T_u$, $T_c$, $T_s$ and $T_0$ in red. Bottom: Computed contribution of FE noise waves for antenna 252A. \label{fig:fee-noisewave}}
\end{figure}

As discussed in Sec.~\ref{sec:calibration},  calibration requires that the noise waves emitted by the receiver are accounted for   (Eq.~\ref{eq:3ss}). Characterization of the receiver's emitted noise wave \citep[see][]{meys1978} requires multiple measurements of output power with varying impedances at the receiver's input. A convenient method for characterization of the noise wave is by applying Eq.~\ref{eq:3ss} to a system where the antenna is replaced by an open (or shorted) coaxial cable; the wrapping of phase $\phi$ as a function of frequency over a cable of suitable length allows deduction of the phase of the emitted noise wave. The process we employed was as follows:

\begin{itemize}
	\item The reflection coefficient $\ss{\Gamma}{coax}$ of an open coaxial cable at room temperature was measured using a VNA, along with $\ss{\Gamma}{rx}$.
	\item The coaxial cable was connected to the FE at the MS147 test port, and power spectra for $\ss{P}{coax}$, $\ss{P}{hot}$ and $\ss{P}{cold}$ were measured using a spectrum analyzer.
	\item A three-state calibrated spectrum was computed via application of Eq.~\ref{eq:3ss} (top panel of Fig.~\ref{fig:fee-noisewave}). 
	\item Replacing the terms $\ss{\Gamma}{ant}$ and $\ss{H}{ant}$ with $\ss{\Gamma}{coax}$ and $\ss{H}{coax}$, we applied least-squares fitting to estimate scalar values $T_c$, $T_s$, $T_u$ and $\psi$. 
\end{itemize}

For the FE corresponding to antenna 252A, we measure $T_U=194.67$, $T_c=-174.39$, and $T_s =-1.14$. The overall magnitude of the noise wave from the receiver, when connected to antenna 252A, is shown in the bottom panel of Fig.~\ref{fig:fee-noisewave} and is at a level of a few percent of the sky temperature.

\subsection{Noise diode thermal stability} \label{sub:diode-stability}

%%%% FIGURE
\begin{figure}
\includegraphics[width=1\columnwidth]{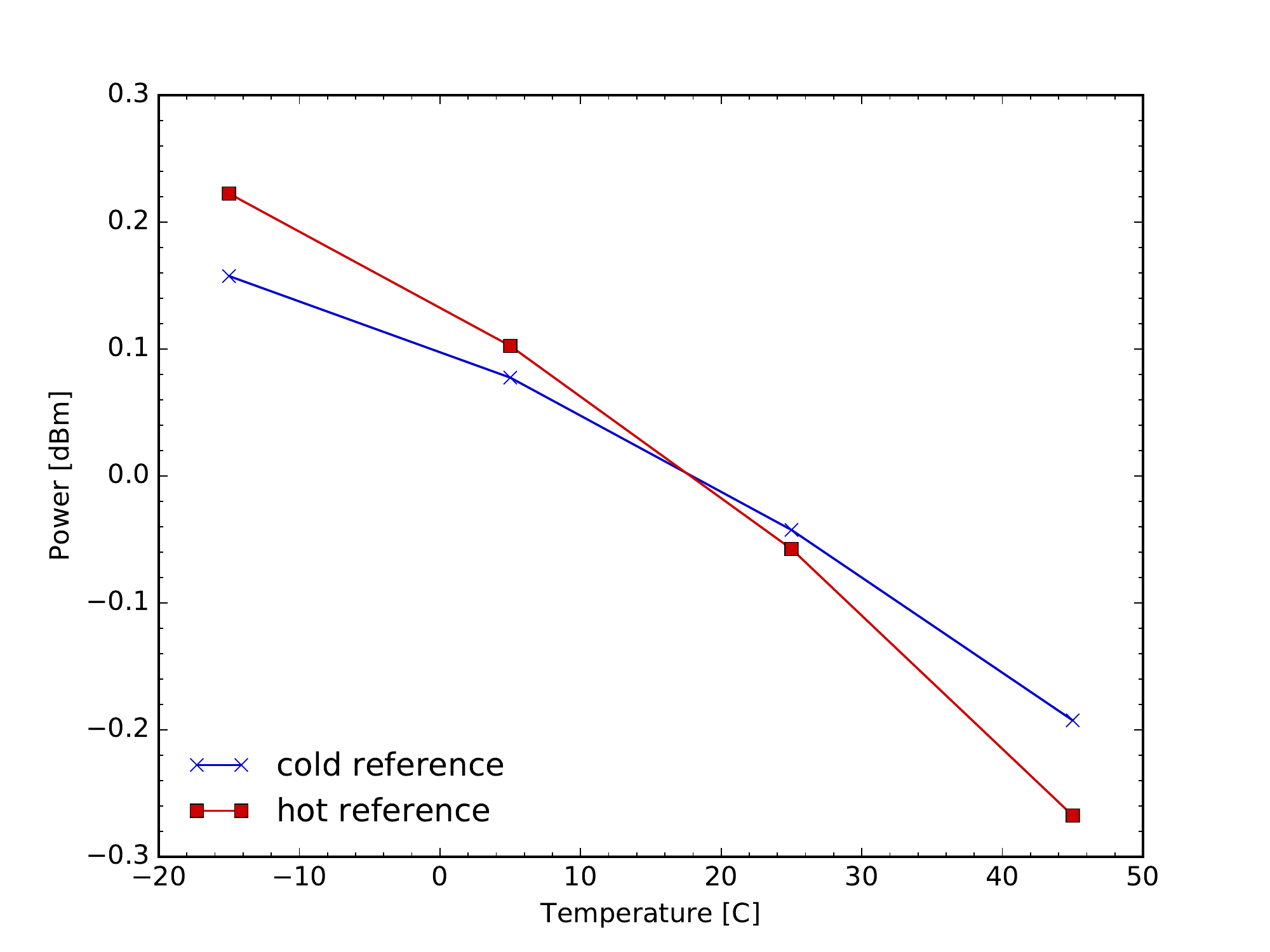}
\protect\caption{Change in noise diode output power as a function of temperature, as measured over hot (red) and cold (blue) reference paths.\label{fig:oven}}
\end{figure}

The characteristics of many active RF devices, including noise diodes, are dependent upon ambient temperature. We used a thermal control chamber (Test Equity 1000) to characterize the effect of ambient temperature upon the output power of the FE noise diode. The chamber allows control of the ambient temperature from $-15^\circ$ to $45^\circ$\,C, with $\pm0.1^\circ$\,C precision.

The FE was placed in the chamber inside an RF-shielded box; a coaxial cable connected the FE output to an HP436A power meter via a bias tee that supplied DC voltage to the FE. The power meter also outputs a 0\,dBm 50\,MHz reference tone; this was connected to the FE input MS147 port, with 60\,dB of attenuation was added at the reference output. We waited 20 minutes between temperature changes, to allow the FE and RF-tight box to equilibrate. 

We found a temperature coefficient of $-0.00815$\,dB/K for the hot reference path, and $-0.00585$\,dB/K for the cold path (Fig.~\ref{fig:oven}). The fractional stability of the two paths is 0.00874\%; as such, temperature dependence of the noise diode is not expected to be a significant source of error.

\subsection{Allan deviation}

%%%% FIGURE
\begin{figure}
\includegraphics[width=1\columnwidth]{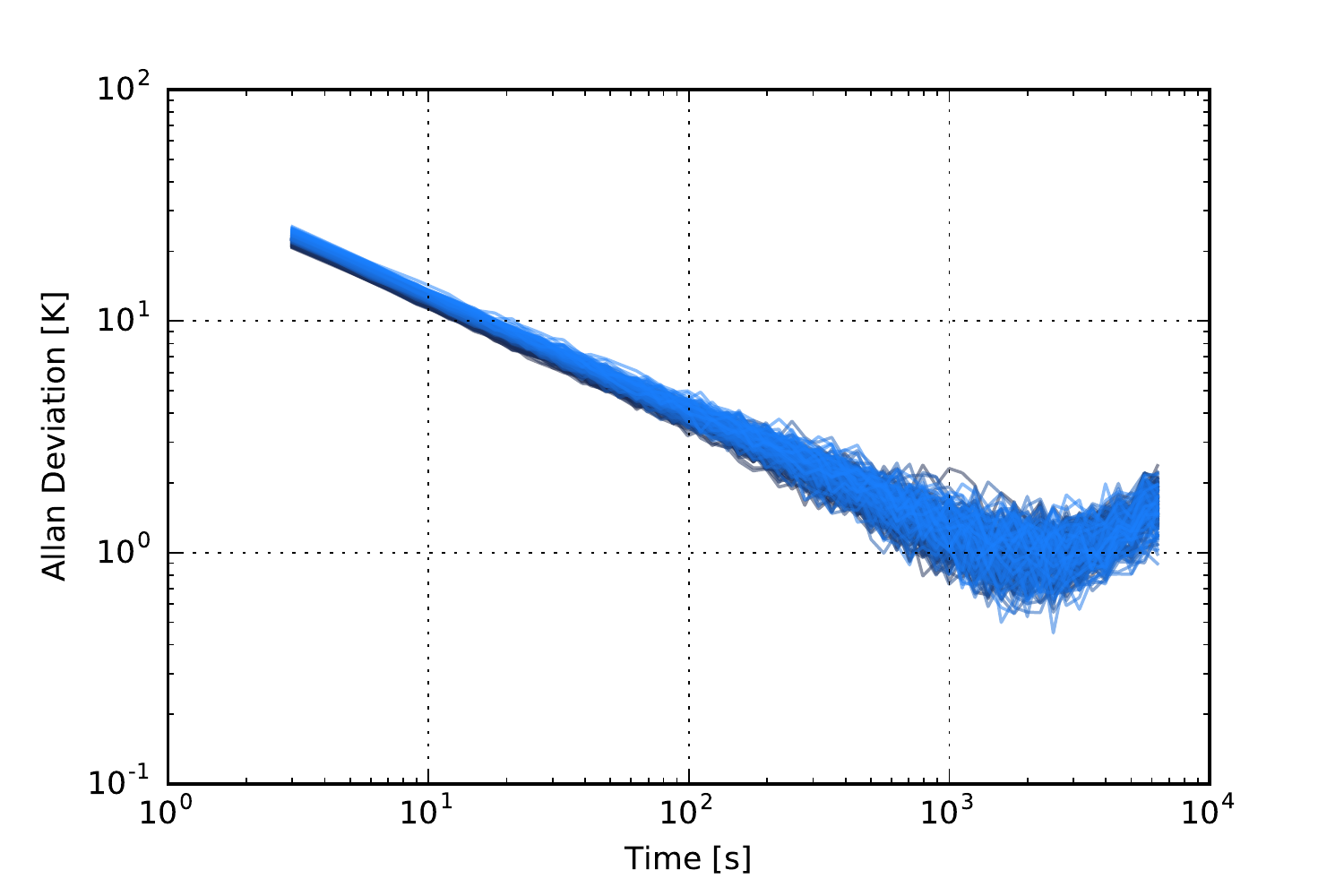}
\protect\caption{Allan deviation for the receiver system connected to a 50\,$\Omega$
load in place of the antenna. Every 7th spectrometer channel is plotted ($\Delta\nu$=24\,kHz) between 30 and 80\,MHz. Data were in the field, with the receiver board installed at the antenna.\label{fig:allan} }

\end{figure}

The Allan variation, $\sigma_y^2(\tau)$, and the Allan deviation, $\sigma_y(\tau)=\sqrt{\sigma_y^2(\tau)}$, are common measures of stability over time. Allan deviation may be used to differentiate between different types of noise within a system. Notably, for random Gaussian noise, such as that in a radiometer, the Allan deviation will decrease as $\tau^{-1/2}$. 

To characterize the stability of the LEDA radiometer system in the field, we took spectrometer data overnight with a 50\,$\Omega$ load connected to the FE MS147 input port. Data were calibrated using the three-state switching method outlined above. We used the \textsc{allantools} package \citep{ascl_allantools} to compute the Allan deviation for the calibrated overnight data (Fig.~\ref{fig:allan}), finding a maximum integration time of $\tau=2000$\,s, before other system instabilities become significant.

\newstuff{By Eqn.~4, for $\Delta\nu$=1\,MHz and sky temperatures of 5000 K, 3000 K and 1000 K, the corresponding rms noise levels are 290\,mK, 174\,mK, and 58\,mK, respectively. As such, the radiometer is stable enough to reach the level required for validation of the \citet{Bowman:2018} result. Nevertheless, as will be discussed later, other systematics currently dominate the noise budget.}  We believe the main source of instability in this field test is the change in the load's ambient temperature overnight, and that the intrinsic stability is higher than that presented here; thermal isolation of the load will be required for future tests.

\section{Results}
\label{Sec:results}

Three LEDA FE boards, as described in Sec.~\ref{sec:calibration}, were deployed at OVRO-LWA in January 2016, to antennas 252, 254 and 255 (Fig.~\ref{fig:ant-pos}). On 2016-01-27, on-sky data were recorded for 24-hours using the LEDA digital spectrometer systems (Sec.~\ref{sec:corr}). For these observations, we switched between the sky and the reference diode states every 5 seconds. For reproducibility, these data, along with analysis scripts used to generate plots in this paper are available online\footnote{\url{http://github.com/telegraphic/leda_analysis_2016}}. 

During the January 2016 deployment, the `B' polarization of board 252 was found to have poor characteristics, so have been excluded from analysis here. In this section, we first present detailed results from a single antenna, before comparing results across antennas in Sec.~\ref{sec:compare-ants}.
 
\subsection{Absolute calibration}
 
\begin{figure*}
\centering
\includegraphics[width=2.0\columnwidth]{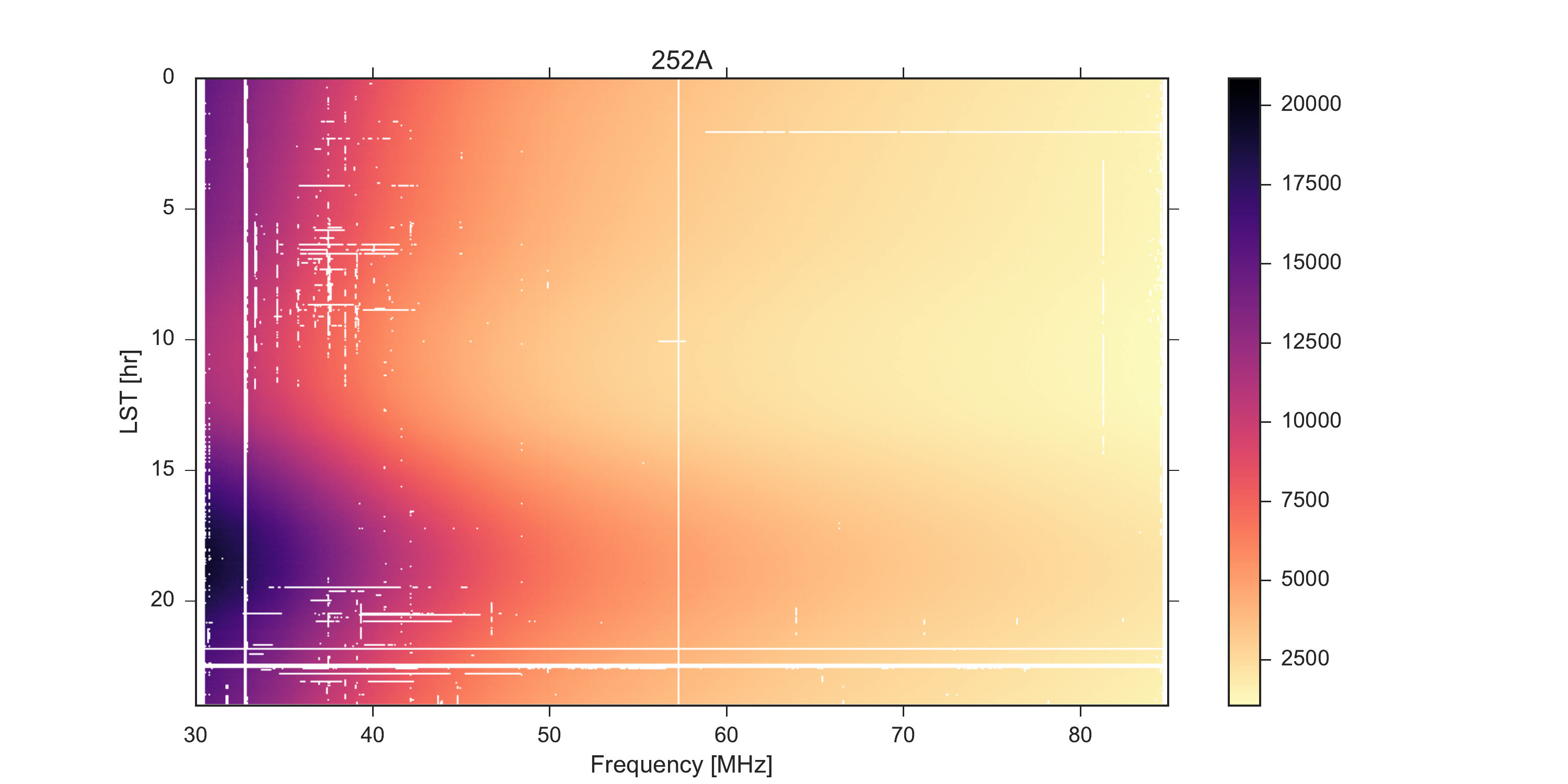}
\protect\caption{ 
	Dynamic spectra for antenna 252A on 2016-01-26 after RFI flagging.  Flagged data are shown in white; color mapping is in Kelvin.
	\label{fig:rfi-flagged}
}
\end{figure*}

\begin{figure}
\centering
\includegraphics[width=1.0\columnwidth]{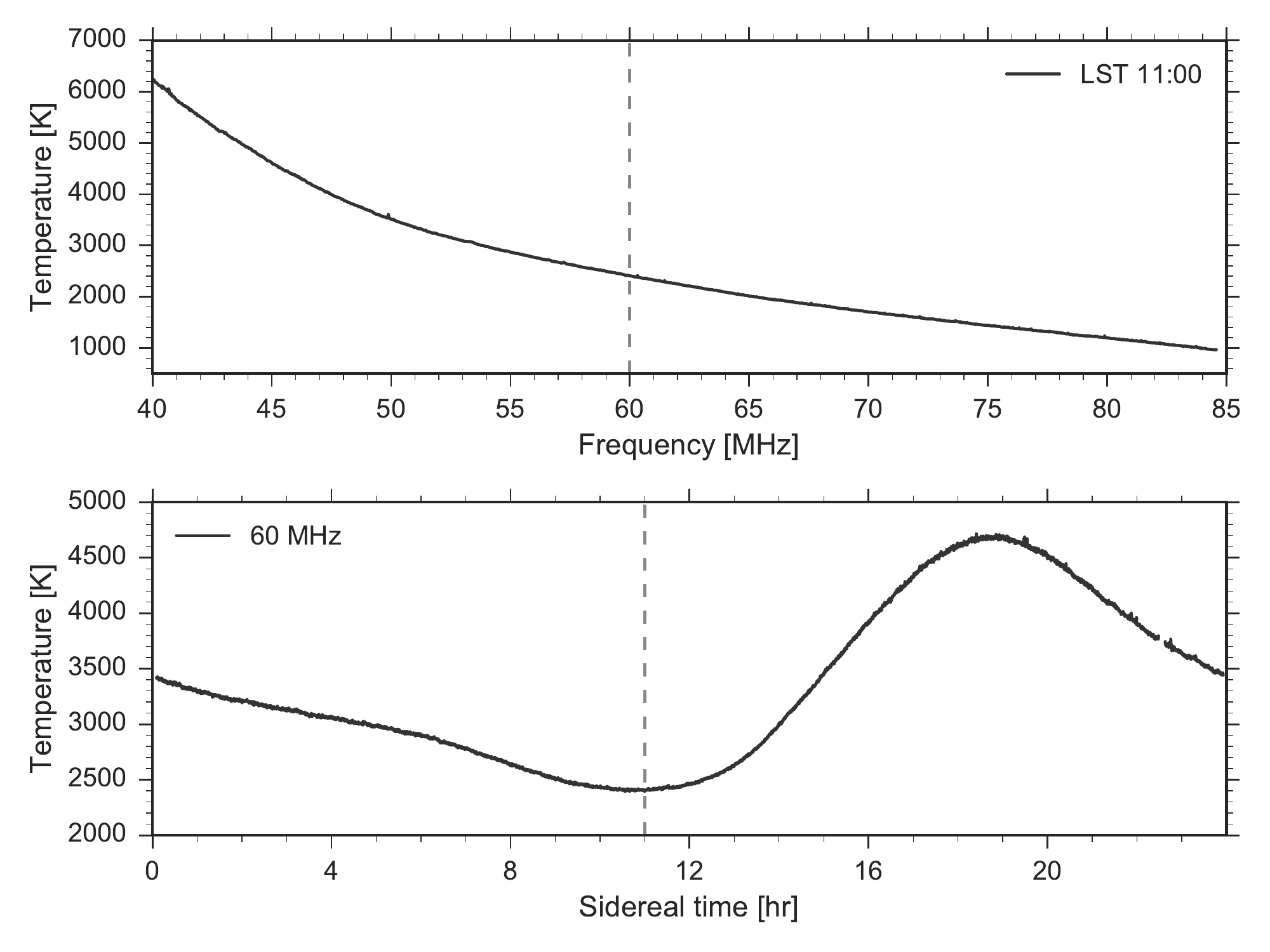}
\protect\caption{ 
	Top: measured antenna temperature for antenna 252A at LST 11:00, 2016-01-26. Bottom: measured antenna temperature at 60\,MHz over a 24-hour period.
\label{fig:a252-spectrum}
}
\end{figure}
 
Data were calibrated using Eq.~\ref{eq:T_3ss}, following the measurement procedures outlined in Sec.~\ref{sec:char}. Fig.~\ref{fig:rfi-flagged} shows the dynamic spectra for antenna 252A over the 24-hour period on 2016-01-26, after RFI flagging (see Sec.~\ref{sec:rfi}). The corresponding antenna temperature spectrum at LST 11:00, is shown in the top panel of Fig.~\ref{fig:a252-spectrum}; the bottom panel shows the change in system temperature over the 24-hour period at 60 MHz. 

\subsection{RFI environment}\label{sec:rfi}

\begin{figure}
\centering
\includegraphics[width=1.0\columnwidth]{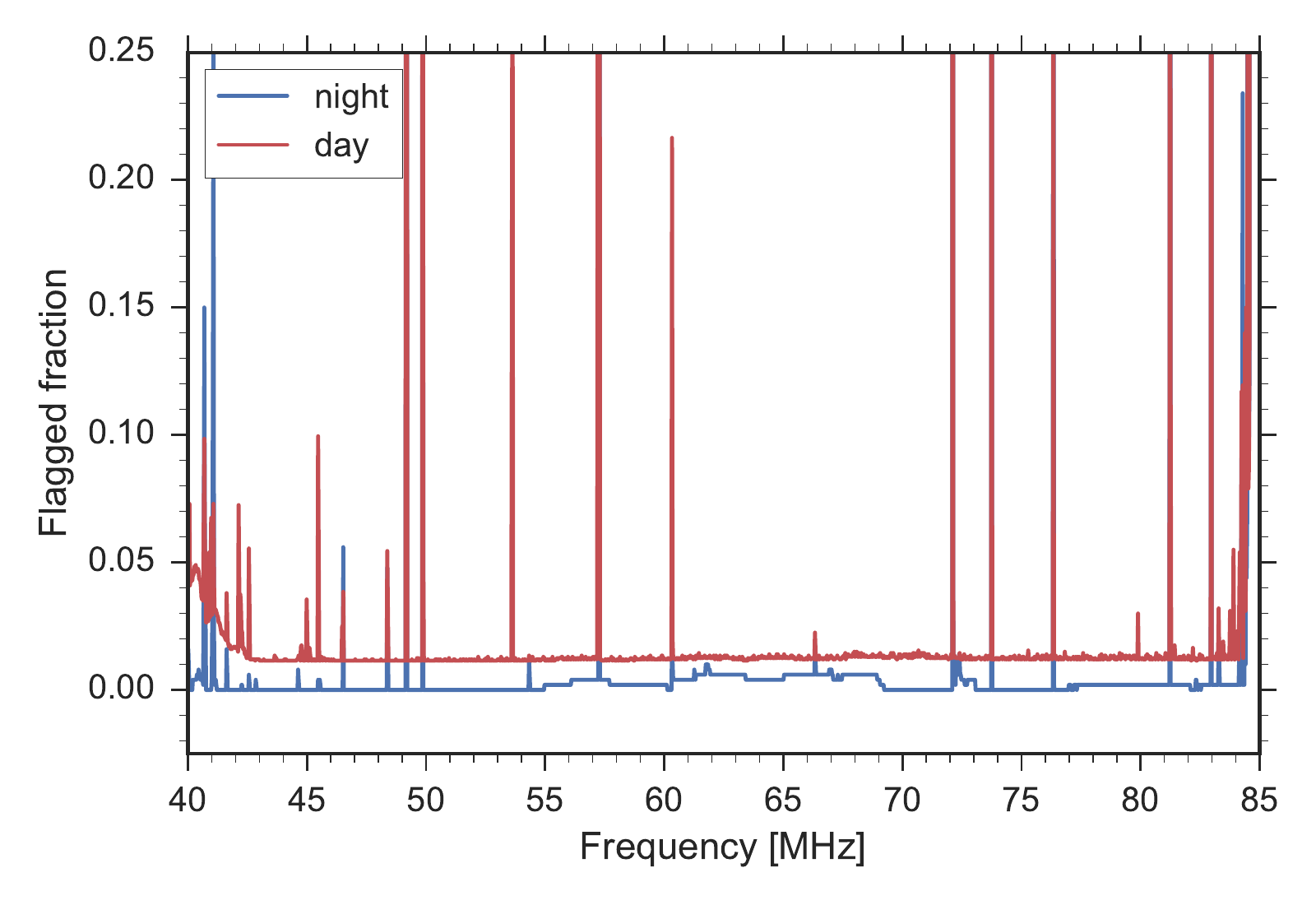}
\protect\caption{ 
	Fraction of data flagged by \textsc{dpflgr} for day (red) and night (blue).
\label{fig:rfi-fraction}
}
\end{figure}

To identify and flag RFI, we apply the \textsc{sumthreshold} algorithm \citep{Offringa2010}, which we have ported to a Python package called \textsc{dpflgr}. Dynamic spectra from antenna 252A post-flagging are shown in Fig.~\ref{fig:rfi-flagged}; the flagged data fractions for day and night are shown in Fig.~\ref{fig:rfi-fraction}. The RFI environment is seen to be quieter at night, but nevertheless several bright narrowband sources are omnipresent. We choose to completely flag channels or timesteps with high occupancy (>40\%). The presence of increased RFI during the day, along with increased air traffic, on-site human activity, and potential solar flare events, motivate primary LEDA observations to be conducted at night.

\subsection{Comparison to sky models}\label{sec:sky-comp}

\begin{figure}
\centering
\includegraphics[width=1.0\columnwidth]{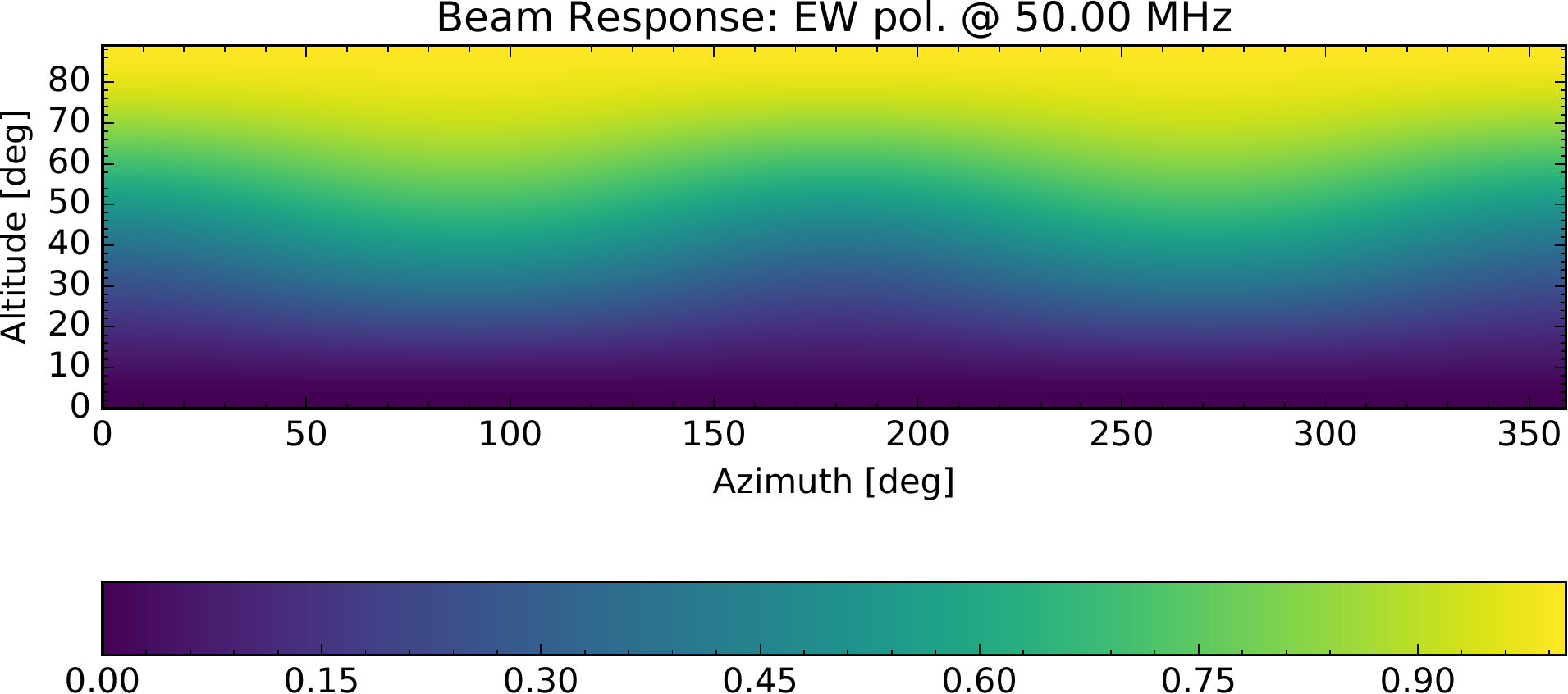}
\protect\caption{ 
	Empirical model of the LWA dipole antenna pattern at 50\,MHz for the East-West polarization.\citep{dowell2017}.  
\label{fig:antenna-pattern}
}
\end{figure}

To compare against our measurements, we simulated the expected antenna temperature spectra using a model of the LWA antenna  with several sky models: the aforementioned GSM \citep[GSM2008,][]{deOliveiraCosta:2008ks}, the `updated' GSM released in 2016 \citep[GSM2016,][]{zheng2017}, and the Low-frequency Sky Model \citep[LFSM,][]{dowell2017}. For the antenna gain pattern, we used an empirical model (valid between 40--80\,MHz) based on LWA1 data \citep{dowell2017}; the response at 50\,MHz is shown in Fig.~\ref{fig:antenna-pattern}.

\begin{figure}
\centering
\includegraphics[width=1.0\columnwidth]{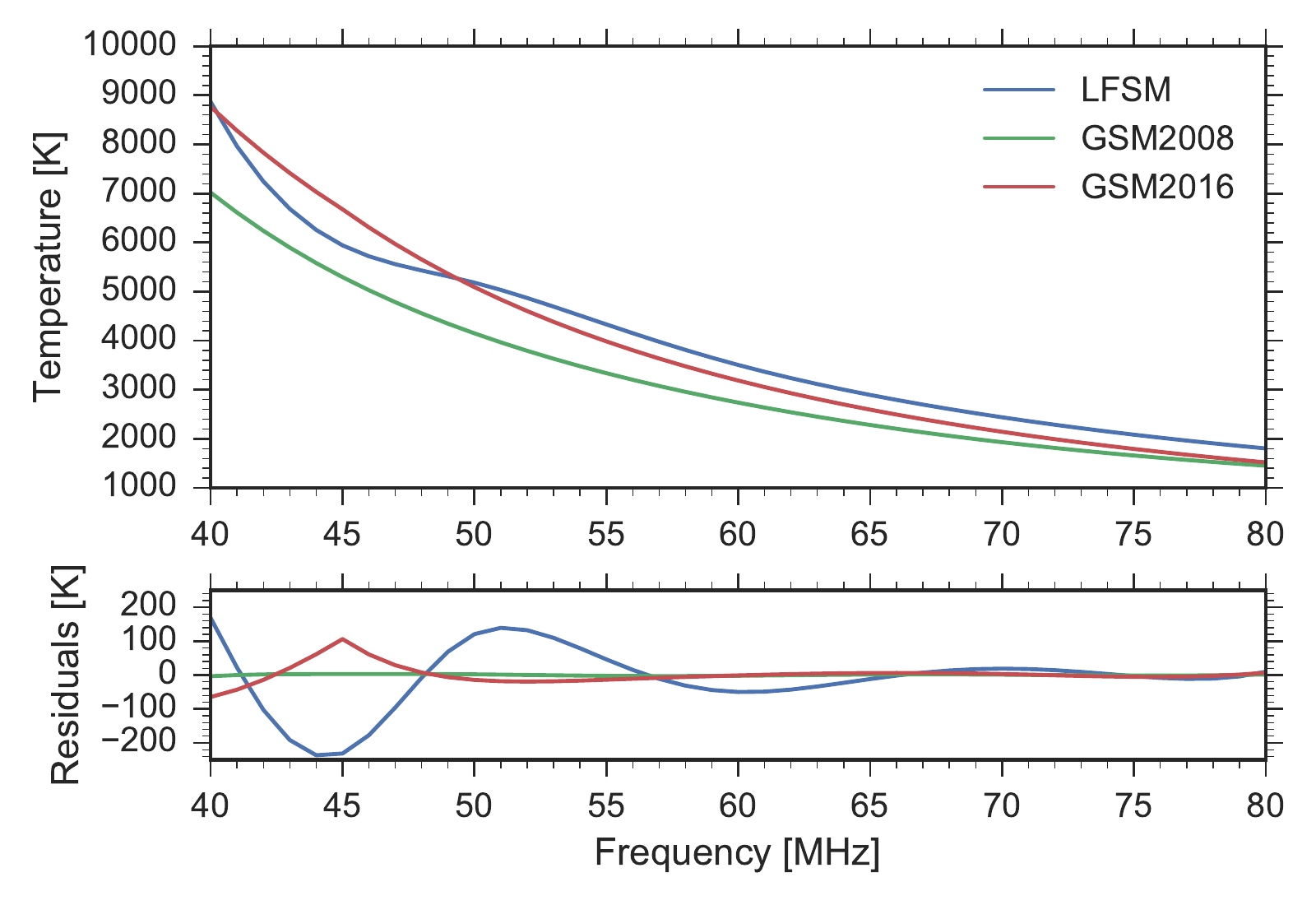}
\protect\caption{ 
	Simulated sky brightness for models of diffuse emission at LST 12:00 (top). A model of the LWA1 antenna is convolved with the sky models for an observer at OVRO-LWA. Bottom: Residuals after subtraction of a 5th order log-polynomial fit. 
\label{fig:skymodels}
}
\end{figure}

Simulated antenna temperature spectra for the three models are shown in Fig.~\ref{fig:skymodels}, for an observer at OVRO, LST 12:00. While the models are in agreement to the 10\% level, the LFSM exhibits an unexpected dip at $\sim$45\,MHz. After subtraction of a 5th order polynomial in log$\nu$ (bottom panel), a discontinuity can be seen in the GSM2016 residual data. The residuals for both the GSM2016 and LFSM are of order $\sim 100$\,K, notably larger than the GSM. The behavior can be traced to the inclusion of data from the \citet{Alvarez1997} 45-MHz survey, suggesting a systematic offset in the underlying data from which the sky model is generated. Due to the unexpected discrepancies in the GSM2016 and LFSM data, we use the GSM2008 as our reference model. 

\begin{figure}
\centering
\includegraphics[width=1.0\columnwidth]{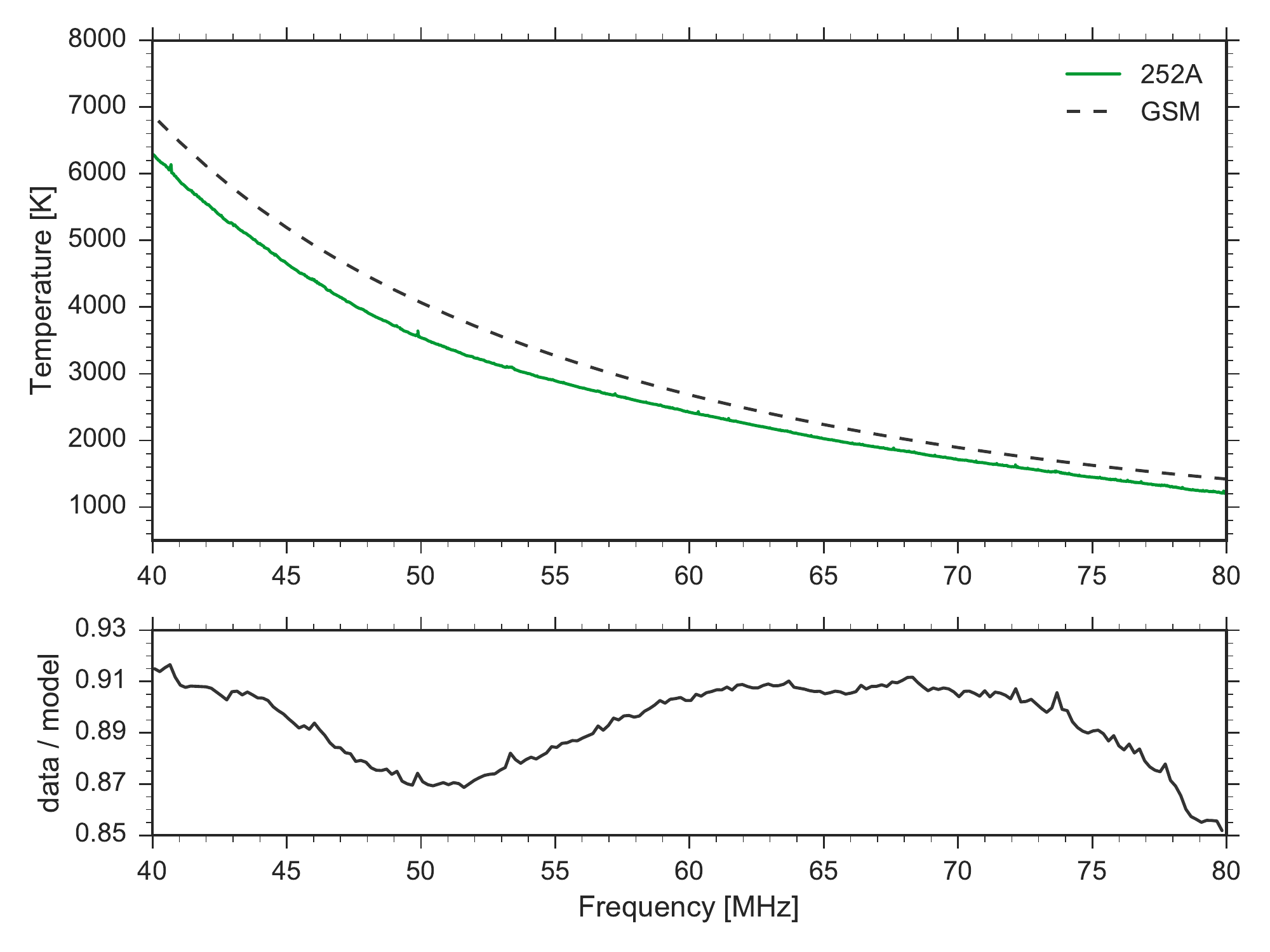}
\protect\caption{ 
	Comparison of simulated data (GSM2008 sky model with LWA antenna model) to calibrated data from antenna 252A, LST 11:00.
\label{fig:skymodel-compare}
}
\end{figure}

A comparison between calibrated spectrum and that expected from the GSM2008 is shown in Fig.~\ref{fig:skymodel-compare}. The ratio between data and model lies between 0.85--0.92 across the 40--80\,MHz band. 

\subsection{Spectral index}

\begin{figure}
\centering
\includegraphics[width=1.0\columnwidth]{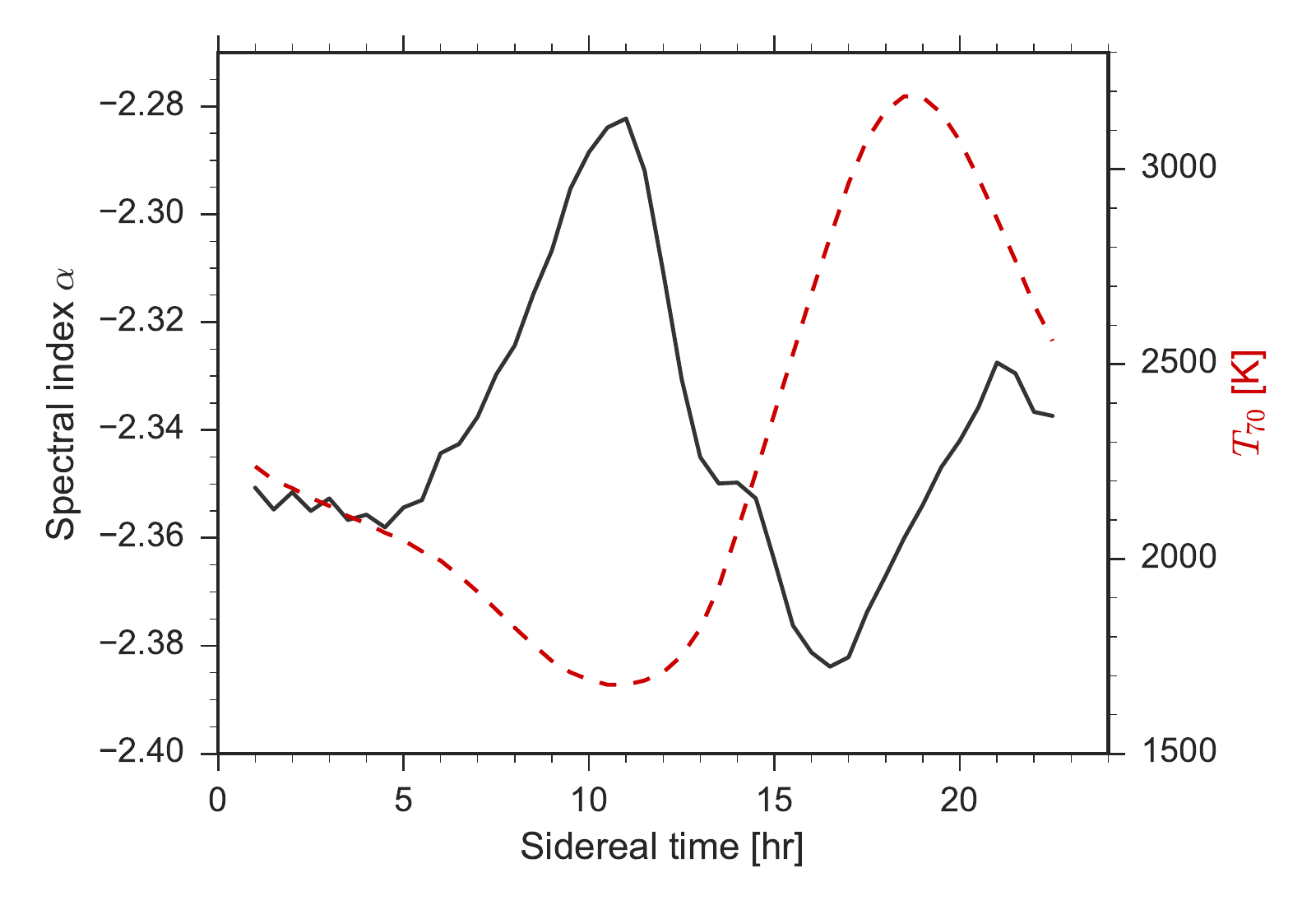}
\protect\caption{ 
	Best fit parameters for spectral index $\alpha$ and $T_{70}$ as a function of LST.  
\label{fig:spec-index}
}
\end{figure}

To compute the spectral index $\alpha$, we perform a least-squares minimization on
\begin{equation}
\chi^{2}=\sum_{i}^{N}\frac{[T_{i}^{{\rm meas}}-T_{70}(\frac{\nu_{i}}{\rm{70 MHz}})^{\alpha} ]^{2}}{\sigma_{i}^{2}}
\end{equation}
where $T^{\rm{meas}}_i$ are our measured sky temperature data per frequency channel $\nu_i$, and $\sigma_i^2$ are per-channel estimates of the thermal noise. We used the \textsc{lmfit} Python package perform the minimization of $\chi^2$ over fit parameters $T_{70}$ and $\alpha$. We find the spectral index varies between $-2.28$ to $-2.38$ over LST (Fig~\ref{fig:spec-index}). These values are consistent with other Northern hemisphere experiments (Tab.~\ref{tab:alpha}). The effect of beam chromaticity \citep{mozdzen2017} is not considered here, and is left for future work.

\begin{table}
	\caption{Experimental measurements of the spectral index of radio emission $\alpha$ below 200 MHz.
	\label{tab:alpha}}
	\begin{tabular}{l c c c}
	\hline
	Reference & Decl. & Freq. & $\alpha$ \\
	          & (deg) & (MHz)     & \\
	\hline
	\hline
	\cite{Costain1960}   &  +52.16   & 38--178   & $-$2.37$\pm$0.04 \\ 
	\cite{Purton1966}    &           & 13--100   & $-$2.38$\pm$0.05 \\
	\cite{Andrew1966}    &  +52.16   & 10--38    & $-$2.43 $\pm$ 0.03 \\
	\cite{Rogers2008}    & $-$26.5     & 100--200  & $-$2.5 $\pm$ 0.1 \\
	\cite{patra2015}     & +13.6     & 110--175  & $-$2.30 to $-$2.45  \\
    \cite{mozdzen2017}   & $-$26.7     & 90--190  & $-$2.5 to $-$2.6 \\
    \emph{This work}     & +37.24    & 40--80   & $-$2.28 to $-$2.38 \\
	\hline
	\end{tabular}

\end{table}

\subsection{Comparisons across antennas \label{sec:compare-ants}}

The fractional difference between spectra integrated for 20 minutes around LST 12:00 are shown in Fig.~\ref{fig:frac-diff}; measurements are consistent to $\pm$5$\%$ between 40--83\,MHz. Above $\sim$83\,MHz, the attenuation due to bandpass filters gives rise to non-linear ADC gain effects, which act to artificially attenuate the sky temperature \citep{Backer2007}. \newstuff{As shown in Fig.~\ref{fig:fee-rx}, the receiver temperature also increases out-of-band. Improving the response above 83\,MHz is an ongoing effort toward verification of the purported \citep{Bowman:2018} absorption feature.}

\begin{figure*}
\centering
\includegraphics[width=2.0\columnwidth]{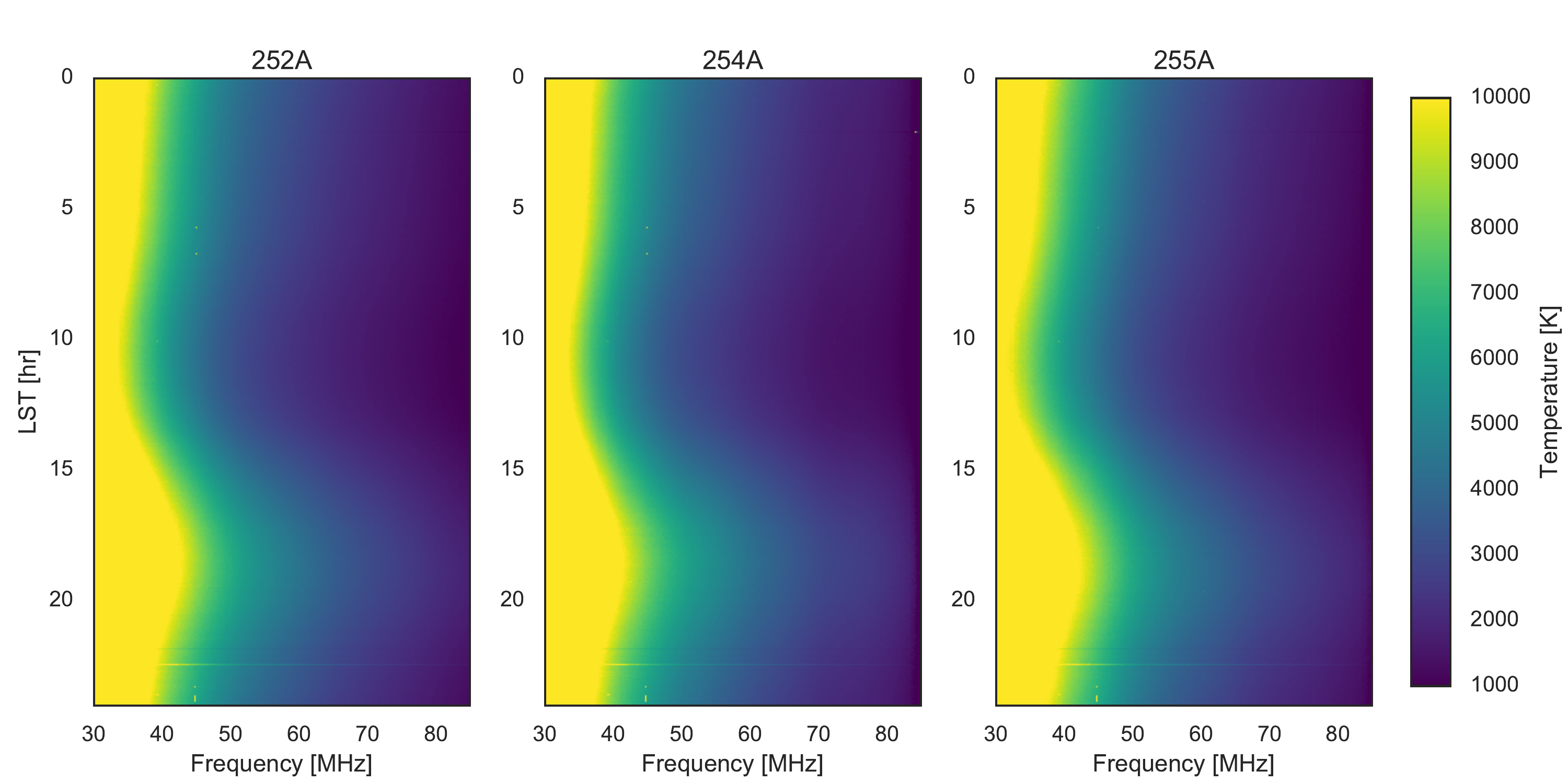}
\protect\caption{ 
	Calibrated sky temperature OVRO-LWA on 2016-01-27, as measured with three independent radiometer systems. Each dynamic spectra shows the sky temperature as a function of frequency over 24 hours of local sidereal time.
	\label{fig:waterfall}
}
\end{figure*}

\begin{figure}
\centering
\includegraphics[width=1.0\columnwidth]{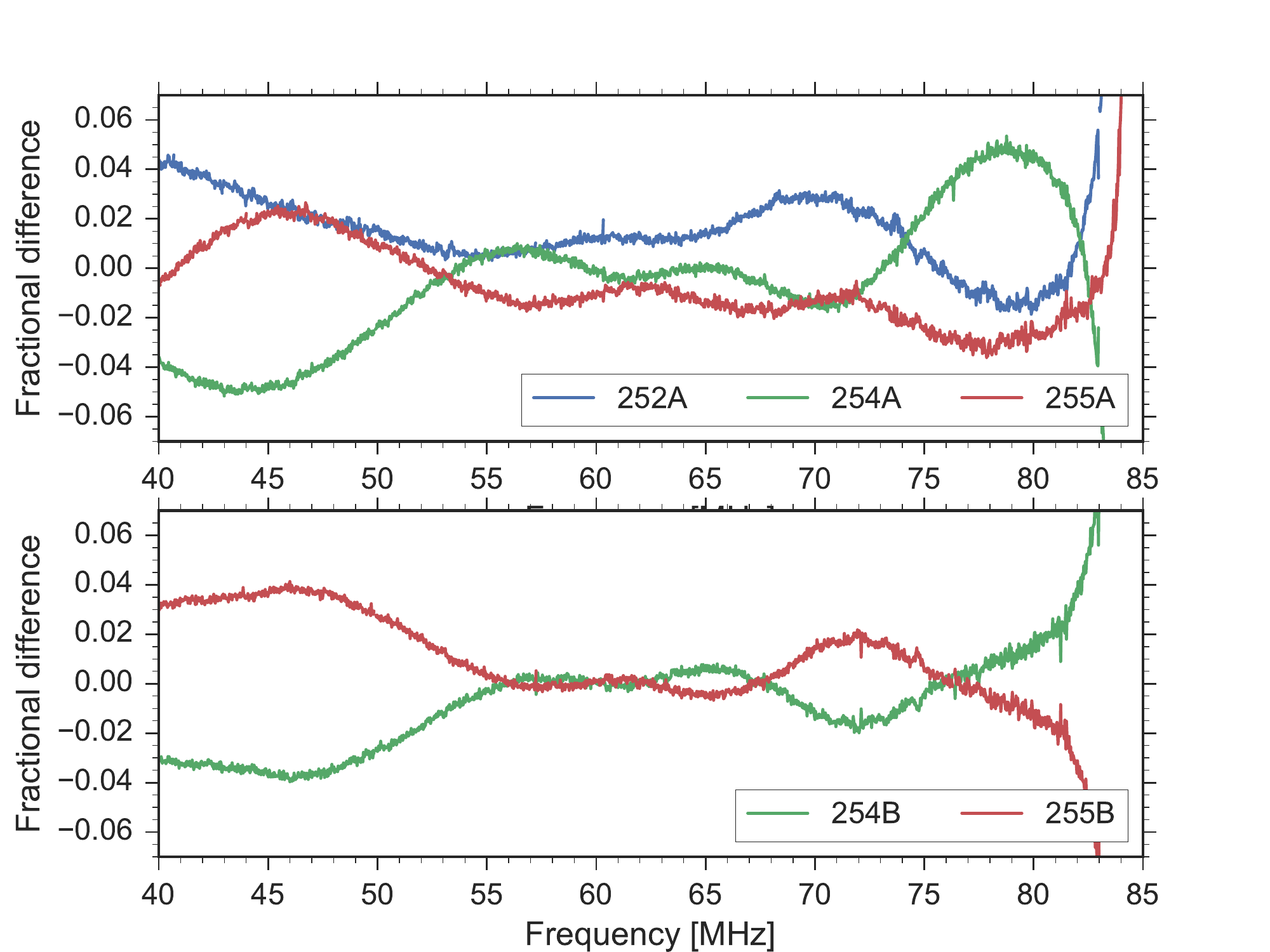}
\protect\caption{ 
Fractional difference between calibrated spectra as shown in Fig.~\ref{fig:waterfall}.
\label{fig:frac-diff}
}
\end{figure}

\begin{figure*}
\includegraphics[width=1.7\columnwidth]{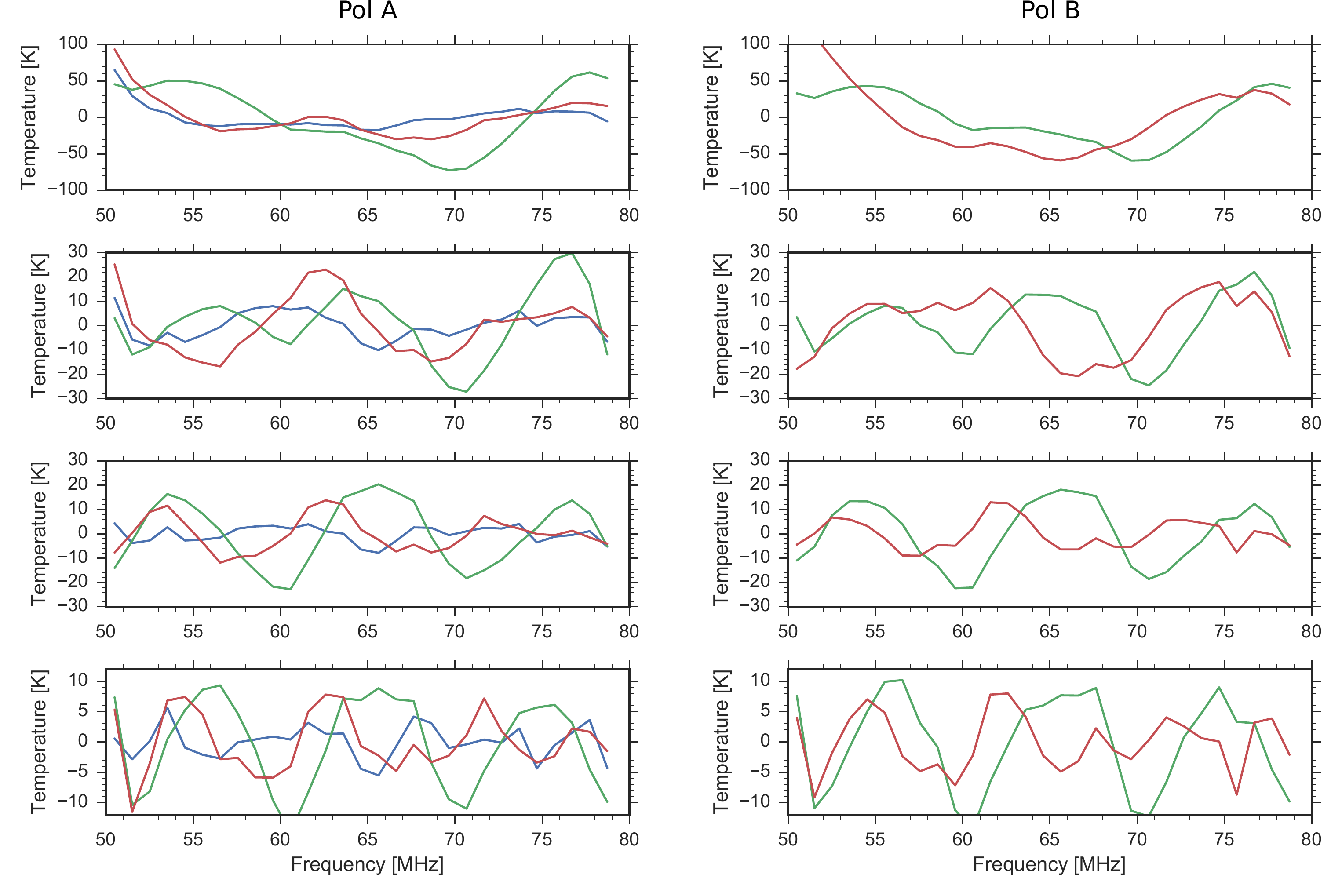}

\caption{Top panel: residuals for calibrated data after subtraction of 1, 3, 5 and 7-term log-polynomial fits, for antennas 252 (blue), 254 (green) and 255 (red); polarization A is shown on the left and polarization B on the right. \label{fig:residuals}}
\end{figure*}

\subsection{Residuals across antennas}

As detailed in \cite{bernardi2015} and \cite{mozdzen2016}, beam chromaticity must be accounted for to mitigate frequency-dependent structure introduced to the global signal. The frequency-dependent response of the antenna must therefore be either simulated using EM software packages such as HFSS and FEKO, or measured directly; LEDA employs the latter approach. {\emph{In-situ}} measurement of the gain pattern of LEDA antennas via cross-correlation with the OVRO-LWA core antennas is beyond the scope of this paper and will be detailed in a future publication. 

Nevertheless, it is illustrative to subtract a log-polynomial sky model from the calibrated data to produce residuals. Fig.~\ref{fig:residuals} shows the residuals after subtraction of log-polynomial fits for LEDA data between 50--80\,MHz, averaged over a one-hour observation period centered an LST of 11:00, 2016-01-26. The calibration and reduction procedure was as follows. Data were calibrated following the absolute calibration approach of Sec.~\ref{sec:calibration}, after which RFI events were flagged (Sec.~\ref{sec:rfi}). After flagging, data were averaged in time (1 hr total) and frequency (1.008 MHz bins) to form mean observed spectra, $T^{{\rm meas}}(\nu)$ for each. 

From top to bottom panel, Fig.~\ref{fig:residuals} shows the residuals of calibrated data after subtraction of 1, 3, 5 and 7-term polynomial fits. We attribute differences between antenna stands primarily to beam chromaticity due to differences in surrounding terrain and differences in as-built antenna geometries. Antenna 252A exhibits the best performance (between $-5$ to 5 K after 7-term fit), with antennas 255A and 255B exhibiting notably higher residual values.

\section{Discussion}
\label{Sec:discussion}

In this paper, we have presented the design and preliminary characterization results for the LEDA radiometer systems. The path toward detection of the 21-cm CD trough will require iterative improvements of the analog systems and analysis methods, as knowledge of the instrumental systematics improve. 

By comparison to the GSM2008, we find our antenna temperature is within $10-15$\% of that predicted for an empirical model of the LWA antenna. Unaccounted for losses in the antenna, or inaccuracy of the maunfacturer-supplied specifications of the HP346C noise source could account for this;  if a multiplicative scale factor of 1.12 is applied, measured data agree with the model to within $\pm3\%$.

We measure the spectral index of the sky to vary between $-2.28$ (LST 11:00) to $-2.38$ (LST 17:00, when the galaxy is high). While in agreement with other observations, we note that beam chromaticity has not been accounted for, which would improve the measurement. Pickup from the ground due to an imperfect ground screen, and from the Sierra Nevada mountain range on the horizon, potentially flatten the true spectral index of the radio sky. Improved measurement of the spectral index is the subject of future work.

\subsection{{\emph{In-situ}} beam measurements\label{sec:beam-meas}}

An important outstanding step is empirical measurement of the antenna gain pattern for each outrigger antenna. As discussed in \citet{bernardi2015} and \citet{mozdzen2016}, gain-pattern-induced chromaticity limits foreground subtraction: this motivated a complete redesign of the EDGES antenna to improve chromatic performance. While an empirical beam model for (closely-packed) LWA antennas has been derived by \citet{dowell2017}, calibration requirements motivate a per-antenna model. Measurements between three antennas (Sec.~\ref{sec:compare-ants}) show agreement to within $\pm$5$\%$ between $40-83$\,MHz; variation in antenna gain pattern may account for much of this. Indeed, these data highlight the advantage that the redundancy offered by multiple measurements of the sky with different radiometer systems provides.

\subsection{Future improvements}

We identify several areas in which our instrument characterization can be improved. Firstly, periodic measurement of scattering parameters measured in the field would allow longitudinal monitoring of the receiver and antenna's reflection coefficients. Additionally, the emitted noise waves could also be measured in the field. Measurement of the noise waves using a commercial impedance tuner in lieu of an open cable would offer an alternative characterization of the LNA noise waves.

Here, we applied the calibration formalism of \citet{rogers2012}. Other approaches, such as the matrix-based calibration approach of \citet{king2010, king2014} offer an alternative approach based on more modern formalisms of noise characteristics. The approach of \citet{Monsalve2017a}, in which extra calibration parameters are included to better fit the data, also offers an alternative avenue toward improved instrument modeling.

Our absolute temperature calibration relies upon an HP346C noise source with manufacturer-supplied characterization. Cross-calibration with other calibration standards, and/or experimental verification of the manufacturer-supplied parameters, may provide improved accuracy of the absolute temperature scaling.

\subsection{Validation of EDGES absorption feature}

\newstuff{Validation of the absorption feature reported by \citet{Bowman:2018} is pressing. As reported here, the LEDA systems exhibit the required radiometric stability, but other systematics, namely the direction-dependent gain of the antennas, confound measurement. Further characterization work is ongoing. }

Upgrades to the LEDA systems are made on a rolling basis. Since the January 2016 campaign---as detailed here---several upgrades have been made to the LEDA systems. These improvements will be discussed further in a future paper. Briefly, radiometric receivers have been installed on all 5 outrigger antennas, modifications to further improve the stability of the noise diode have been made, and a logging system for measurement of the ambient temperature at the antenna has been added. \newstuff{Of importance to validation of the \citet{Bowman:2018} signal, bandstop filters with sharper roll-off have been sourced to allow access to frequencies of up to 87.5\,MHz, while still strongly attenuating the 88--108\,MHz FM band.}

An observation campaign with the upgraded LEDA system was undertaken over November 2016--March 2017; analysis of these data, along with data from 2018, is ongoing.

\section{Conclusions}
\label{Sec:conclusions}

Measurement of the 21-cm emission from the early Universe via radiometric methods requires exquisite calibration and comprehensive knowledge of the radiometer systems. \newstuff{The purported detection of a 21-cm absorption feature during Cosmic Dawn by \citet{Bowman:2018} suggests that the radiometric approach does indeed offer a window into Cosmic Dawn. Validation of the \citet{Bowman:2018} signal is pressing, particularly given the concerns raised by \citet{Hills:2018}, and is an exciting opportunity for radiometric Cosmic Dawn experiments such as LEDA, SARAS 2 and PRIZM.}

In this paper, we have presented the design and characteristics of the LEDA radiometer systems. Comparison of the system performance with predictions based on the GSM2008 sky model and LWA antenna gain pattern are in agreement to the 15\% level over 40--83\,MHz. Between antennas, data agree to $\pm5\%$. Above 83\,MHz, the rolloff of the filter for FM-band rejection (88--108\,MHz) becomes significant.

\newstuff{Upgrades to increase the LEDA observation window cutoff from 83\,MHz to 87.5\,MHz are underway. Further characterization work is also ongoing, in order to place limits on the 21-cm emission during Cosmic Dawn. In particular, individual characterization of the antenna's direction-dependent gain may be needed to account for the frequency dependence of the beam. Work on this characterization is underway, using interferometric measurements with the combined LEDA radiometer antennas and OVRO-LWA core antennas.}

\section*{Acknowledgments}

This work has benefited from open-source technology shared by the Collaboration for Astronomy Signal Processing and Electronics Research (CASPER). We thank the Xilinx University Program for donations; NVIDIA for proprietary tools, discounts, and donations; Digicom for collaboration on manufacture and testing of samplers and ROACH2 processors; and Y. Belopolsky (Bel-Stewart R\&D) for collaboration in development of CAT-7A ARJ45 pass-through hardware and cable assemblies.  We thank R. Blundell, Ed Tong, and P. Riddle of the Smithsonian Astrophysical Observatory Submillimeter Receiver Lab for collaboration on development and fabrication of receivers and other LEDA signal path and control elements.  The great dedication, innovation, and exemplary skill of the Caltech Owens Valley Radio Observatory staff, demonstrated in constructing the LWA array, having created a purpose-built facility in no time deserves special mention.  LEDA research has been supported in part by NSF grants AST/1106059, PHY/0835713, and OIA/1125087. The OVRO-LWA project was enabled by the kind donation of Deborah Castleman and Harold Rosen. GB acknowledges support from the Royal Society and the Newton Fund under grant NA150184. This work is based on the research supported in part by the National Research Foundation of South Africa under grant 103424. GH acknowledges the support of NSF CAREER award AST-1654815.

\bibliographystyle{mnras}
\bibliography{references}

\bsp

\label{lastpage}
\end{document}